\documentclass[
superscriptaddress,
showpacs,
amssymb,
10pt,
reprint,
aps,
prd,
longbibliography,
nofootinbib,
floatfix
]{revtex4-2}

\usepackage[T1]{fontenc}
\usepackage[utf8]{inputenc}
\usepackage{times}

\usepackage[dvipsnames]{xcolor}

\usepackage{amsmath,amssymb,amsfonts}
\usepackage{mathtools}
\usepackage{bm}
\usepackage{bbm}
\usepackage{mathrsfs}
\usepackage{leftindex}
\usepackage{tensor}
\usepackage{accents}
\usepackage{scalerel}

\usepackage{mhchem}
\usepackage{siunitx}

\usepackage{graphicx}
\usepackage{subfigure}
\usepackage[percent]{overpic}
\usepackage{tikz}
\usetikzlibrary{calc}

\usepackage{booktabs}
\usepackage{dcolumn}
\usepackage{enumitem}

\usepackage{soul}
\usepackage[normalem]{ulem}
\usepackage{comment}

\usepackage{orcidlink}
\usepackage{hyperref}

\definecolor{darkblue}{rgb}{0,0,0.5}
\definecolor{navy}{RGB}{0,0,150}

\hypersetup{
    bookmarks=true,
    pdftoolbar=true,
    pdfmenubar=true,
    pdffitwindow=true,
    pdfstartview={FitH},
    pdftitle={My title},
    pdfauthor={author},
    pdfsubject={Subject},
    pdfcreator={Creator},
    pdfproducer={Producer},
    pdfkeywords={keyword1},
    pdfnewwindow=true,
    colorlinks=true,
    linkcolor=darkblue,
    citecolor=darkblue,
    filecolor=navy,
    urlcolor=darkblue
}

\newcommand{\ed}{\mathrm{d}}
\newcommand{\ba}{\bar{a}}

\newcommand{\R}{\mathcal{R}}

\newcommand{\Q}{\mathscr{Q}}

\begin{document}

\title{Strong-lensing degeneracies of black holes embedded in self-interacting scalar field dark matter halos}

\author{Mohsen Fathi\orcidlink{0000-0002-1602-0722}}
\email{mohsen.fathi@ucentral.cl}
\affiliation{Centro de Investigaci\'{o}n en Ciencias del Espacio y F\'{i}sica Te\'{o}rica (CICEF), Universidad Central de Chile, La Serena 1710164, Chile
}

\author{Gabriel G\'{o}mez\orcidlink{0000-0002-3618-9824}}
\email{luis.gomezd@umayor.cl}
\affiliation{Centro Multidisciplinario de F\'{i}sica, Vicerrector\'{i}a de Investigaci\'{o}n, Universidad Mayor,
Camino La Pir\'{a}mide 5750, Huechuraba, 8580745, Santiago, Chile}

\begin{abstract}

In this paper, we explore the strong gravitational lensing properties of black holes embedded in self-interacting scalar field dark matter halos, together with NFW-type configurations for comparison. The corresponding spacetime geometry is reconstructed numerically through the Einstein cluster formalism, allowing us to study how the surrounding dark matter distribution affects the propagation of photons near the black hole. We first analyze the effective function governing photon trajectories and calculate the corresponding photon sphere radius and critical impact parameter. We then investigate different strong-lensing observables, including relativistic Einstein rings, finite-order image positions, image separations, magnifications, and time delays, with particular attention to the supermassive black holes M87* and Sgr A*. Our results show that the considered halo configurations produce only small deviations with respect to the Schwarzschild case, typically at the level of $\mathcal{O}(10^{-3})$ or smaller, leading to a strong observational degeneracy among the models. Nevertheless, small but systematic differences remain present, especially in the time delay between relativistic images, which provides the clearest amplification of the halo-induced corrections for very massive black holes. These results suggest that, although standard strong-lensing observables remain highly robust against the considered halo environments, time-domain signatures may offer a more promising way to probe the effect of dark matter surrounding black holes.

\bigskip

{\noindent{\textit{keywords}}: black holes, dark matter, strong gravitational lensing, photon sphere, relativistic images}\\


\end{abstract}

\maketitle

\section{Introduction}

Black holes are usually treated first as isolated vacuum solutions of general relativity. The Schwarzschild and Kerr metrics are the basic examples of this description, and they have been extremely successful in the study of compact objects and their relativistic effects~\cite{Schwarzschild:1916,Kerr:1963,Chandrasekhar:579245,Misner:1973,Will:2014kxa}. However, real astrophysical black holes are not completely isolated. They are surrounded by accretion matter, magnetic fields, plasma, stars, and also by the dark matter distribution of the host galaxy. Therefore, even if the vacuum geometry gives the dominant part of the strong-field region, it is natural to ask how much the external environment can change the propagation of light around the black hole.

Dark matter is one of the main ingredients of the present cosmological picture. Its existence is supported by galactic rotation curves, galaxy clusters, gravitational lensing, and the large-scale structure of the Universe~\cite{Rubin:1980,Bertone:2005,Planck:2018vyg,Bullock:2017xww}. The standard cold dark matter (CDM) scenario describes many observations very well, and the Navarro--Frenk--White (NFW) profile is one of the most used halo models in this context~\cite{Navarro:1996,Navarro_2007}. Nevertheless, some tensions at galactic scales, such as the cusp-core problem, have also motivated alternative dark matter models~\cite{deBlok:2009sp}. Among them, scalar-field dark matter, fuzzy dark matter, and self-interacting dark matter have received much attention~\cite{Spergel:1999mh,Hu:2000ke,Schive:2014dra,Hui:2016ltb,Tulin:2017ara}. In particular, self-interacting scalar field dark matter can form a core--halo structure, where the inner solitonic region is connected to an outer CDM-like halo~\cite{Fan:2016rda,Matos:2000ss,Robles:2012uy,UrenaLopez:2019kud}. Such a structure is interesting for black hole physics because the halo modifies the mass distribution around the central object, even if the modification near the photon sphere may be small.

The effect of dark matter around black holes has been studied in several directions. Dark matter can form spikes or dense distributions around massive black holes, and this may affect the motion of stars, compact objects, and photons~\cite{Gondolo:1999ef,Sadeghian:2013laa,Eda:2013gg,Eda:2014kra,Lacroix:2013}. Dark matter environments can also influence gravitational wave signals and the dynamics of compact binaries~\cite{Gomez:2017dhl,Eda:2015dka}. In the same spirit, different black hole metrics surrounded by dark matter or dark fluids have been investigated in connection with geodesics, shadows, thermodynamics, accretion, and lensing~\cite{Li:2023zfl,kiselev_quintessence_2003,Mustafa:2023,Arora:2023,xu_black_2020,konoplya_shadow_2019,jusufi_shadows_2020,li_shadow_2020,pantig_dark_2022,pantig_dehnen_2022,pantig_black_2023,pantig_apparent_2024}. These works show that the surrounding matter sector can leave small but systematic changes in the photon motion. Related studies have also considered weak lensing and shadows in spacetimes with dark matter or other non-vacuum structures~\cite{Ovgun:2018oxk,Ovgun:2018fnk,Ovgun:2019jdo,Ovgun:2020gjz,Ovgun:2021ttv,ovgun_constraints_2024,Yang:2024dmhalo}.

Gravitational lensing is one of the most direct tools to test the geometry around compact objects. In the strong-deflection regime, photons can pass close to the photon sphere, wind several times around the black hole, and then reach the observer as a sequence of relativistic images~\cite{Darwin_gravity_1959,Virbhadra:2000,Bozza:2001,Bozza:2002,Bozza:2010,bozza_strong_2007,Bozza:Scarpetta:2007,Perlick:2022}. The corresponding observables include the limiting angular position $\theta_\infty$, the separation $s$ between the first image and the packed inner images, the relative magnification, and the time delay between different relativistic images~\cite{Bozza:Mancini:2004,Bozza:2004,bozza_time_2004}. In recent years, finite-distance corrections and higher-order images have also been studied in more detail, because realistic sources and observers are not necessarily located at infinity~\cite{bozza_strong_2007,bisnovatyi-kogan_analytical_2022,tsupko_shape_2022,aratore_constraining_2024}. This is important for the present work, since we want to compare not only the standard shadow-size quantities, but also more detailed observables which may respond differently to the halo structure.

The observational motivation has also become stronger after the Event Horizon Telescope observations of M87* and Sgr A*~\cite{Akiyama:2019,Akiyama:2022,the_event_horizon_telescope_collaboration_first_2019,event_horizon_telescope_collaboration_first_2022}. These observations are compatible with the Kerr picture within present uncertainties, but they also open the possibility of testing small deviations coming from modified gravity, plasma effects, or environmental matter~\cite{Psaltis:2019,Gralla:2019xty,Gralla:2019,johnson_universal_2020,Vagnozzi:2022moj,Vagnozzi:2022tba,vagnozzi_horizon-scale_2023}. In this direction, black hole shadows and lensing observables have been used to constrain several non-vacuum or non-standard black hole geometries~\cite{Atamurotov:2013sca,Atamurotov:2015nra,Perlick:2015vta,Perlick:2018a,xavier_shadows_2023}. For dark matter halos specifically, the question is more subtle: the effect is expected to be small close to the photon sphere \cite{Gomez:2026skm}, but it may still produce coherent changes in the critical impact parameter, the relativistic images, or the time-delay scale.

In this paper, we study these effects for black holes embedded in self-interacting scalar field dark matter halos. The geometry is reconstructed numerically by using the Einstein cluster approach, where the matter sector is anisotropic, with vanishing radial pressure and non-zero tangential pressure. This formalism has been recently used to describe black holes in dark matter environments and allows one to obtain the lapse function directly from the Einstein equations~\cite{Figueiredo:2023gas}. We mainly focus on the self-interacting scalar field dark matter profile, but we also include NFW-type configurations for comparison. Similar motivations have appeared in recent works on black hole shadows and dark matter constraints, including the study of self-interacting scalar field dark matter from EHT shadow data~\cite{Gomez:2024sfdm}, as well as earlier works on strong lensing by fermionic dark matter distributions~\cite{Gomez:2016lensing}. These studies show that lensing is a useful way to connect compact-object observables with dark matter distributions.

Our purpose is not only to compute the usual shadow and strong-lensing observables, but also to see which of them is really useful to distinguish between different halo configurations. For this reason, after calculating the photon sphere and the critical impact parameter, we analyze the deflection angle, relativistic Einstein rings, finite-order image positions, image separations, magnifications, and time delays. We apply these quantities to M87* and Sgr A* and compare the different halo models with the Schwarzschild case. In this way, the paper is mainly focused on the observational degeneracy of the lensing signatures and on the hierarchy of sensitivity among the observables.

The paper is organized as follows. In Sec.~\ref{sec:BHDM_sys}, we introduce the black hole--dark matter system and the halo profiles used in the numerical analysis. In Sec.~\ref{sec:ShDe_gen}, we derive the null geodesic equations and identify the photon-sphere condition. In Sec.~\ref{sec:lensing}, we review the finite-distance strong-lensing formalism used in this work. In Sec.~\ref{sec:lensEq}, we compute the standard lensing observables and compare them with EHT-related constraints. In Sec.~\ref{sec:higher_order}, we go beyond the primary observables and analyze finite-order images, magnifications, time delays, and the hierarchy of lensing signatures. Finally, in Sec.~\ref{sec:conclusions}, we summarize the main results of this work and comment on possible future developments.

\section{Black hole-dark matter system}\label{sec:BHDM_sys}

There are several approaches to modeling a black hole embedded in a dark matter halo, each offering a more realistic description of the astrophysical environment surrounding galactic centers. In this work, we focus on the Einstein cluster method, which has been recently employed to study the influence of dark matter halos on gravitational wave emissions \cite{Figueiredo:2023gas}. This method considers test particles distributed over circular geodesic orbits, leading to an anisotropic but spherically symmetric system with tangential pressure and vanishing radial pressure. The energy-momentum tensor for this system is given by
\begin{equation}
    T^{\mu}{}_{\nu} = \mathrm{diag}(-\rho, 0, P_t, P_t).
\end{equation}
 Here $\rho$ and $P_{t}$ are, respectively, the energy density and tangential pressure of the dark matter distribution. The spacetime can be modeled by the line element  
\begin{eqnarray}
    ds^{2}&=&-f(r)dt^{2}+\frac{dr^2}{g(r)}+r^{2}\left( d\theta^2 + \sin^2\theta d\phi^2\right),\nonumber\\
    g(r) &=& 1-\frac{2m(r)}{r},
    \label{eq:metric0}
\end{eqnarray}
in the Schwarzschild coordinates $(t,r,\theta,\phi)$, where $f(r)$ is the lapse function, 
and $m(r)$ is the total mass function. The non-vacuum Einstein equations $G^{\mu}_{\nu}=8\pi T^{\mu}_{\nu}$ read
\begin{equation}
   m^{\prime}(r)=-4\pi r^{2} \rho(r),\label{eq:Einstt}
\end{equation}
\begin{equation}
   \frac{rf^{\prime}(r)}{2f(r)}=\frac{m(r)}{r-2m(r)}.\label{eq:Einsrr}
\end{equation}
The latter relation allows us to construct $f(r)$ for a given mass function $m(r)$, which contains information about both the black hole and the dark matter configuration. From the Bianchi identities, which imply the conservation equation $\nabla_{\mu}T^{\mu\nu}=0$, one obtains a relation for the tangential pressure of the form
\begin{equation}
    P_{t}=\frac{m(r)/2}{r-2m(r)}\rho,
\end{equation}
To fully determine the system, it remains necessary to specify the dark matter distribution
$\rho(r)$. Additionally, a suitable set of boundary conditions must be considered when integrating Eqs.~(\ref{eq:Einstt}) and (\ref{eq:Einsrr}). Specifically, at the horizon radius, $r_{h}=2M_{\rm BH}$, we impose the condition that  the mass function $m(r_{h})=M_{\rm BH}$, where $M_{\rm BH}$ is the central black hole mass. For radii outside the horizon, the dark matter distribution contributes to the mass function. In practice, this contribution is assumed to become relevant only beyond an inner cutoff radius, $r>r_{h}$, typically identified with the marginally
bound orbit radius $r_{\rm mb}=2r_{h}$. From this radius outward, the dark matter density contributes to the mass function up to the halo radius $R$.\footnote{This prescription avoids introducing an \emph{ad hoc} cutoff in the density profile, such as the commonly used rescaling $\rho(r)\to\rho(1-2M_{\rm BH}/r)$, which artificially forces the density to vanish at the horizon.} These boundary conditions can be summarized as follows
\begin{equation}
\label{eq:IC_mass}
m(r)=
\begin{cases}
M_{\rm BH}, & r = r_h, \\[4pt]
M_{\rm BH} + m_{\rm DM}(r), & r_h < r < R, \\[4pt]
M_{\rm BH} + M_{\rm DM}, & r > R.
\end{cases}
\end{equation}
where $m_{\rm DM}(r)$ represents the fraction of dark matter mass surrounding the black hole
\begin{equation}
    m_{\rm DM}(r)=\int_{r_{\rm mb}}^{R} dr\; r^{2}\; \rho(r),
\end{equation}
and $m_{\rm DM}(r>R)=M_{\rm DM}$ denotes the total dark matter mass enclosed by the halo radius $R$. Additionally, we must choose the following boundary conditions for the lapse function:
\begin{equation}
\label{eq:IC_metric}
f(r)=
\begin{cases}
0, & r = r_h, \\[4pt]
1 - \dfrac{2(M_{\rm BH} + M_{\rm DM})}{r}, & r > R.
\end{cases}
\end{equation}
Outside the radius $R$, we recover the vacuum Schwarzschild solution with an effective mass shifted by the contribution of the dark matter halo.  We ensure that the metric remains regular everywhere, provided all of the above conditions are satisfied. Up to this point, we have introduced a formalism to describe the gravitational potential of a Schwarzschild black hole embedded in a generic dark matter environment. Next, we will specify the dark matter distribution in detail.

\subsection{Core--halo structure of self-interacting scalar field dark matter}

In the Thomas--Fermi (TF) regime, self-interacting scalar field dark matter (SI-SFDM) is characterized by a core--envelope halo structure \cite{Goodman:2000tg,Boehmer:2007um}. 
The inner region forms a solitonic core supported by the strong repulsive quartic self-interaction, while the contribution from quantum pressure is negligible. Beyond the soliton radius, both quantum pressure and self-interactions become subdominant, and the extended halo is supported by the velocity dispersion of the dark matter particles, which balances gravitational collapse \cite{Chavanis:2018pkx}.

Within this framework, a convenient semi-analytical model for the SI-SFDM halo can be constructed by matching a soliton core to an outer CDM-like envelope. This structure is motivated by one-dimensional hydrodynamical simulations of halo formation \cite{Dawoodbhoy:2021beb,Garcia:2023abs}. Accordingly, the density profile of the halo can be written as
\begin{equation}
\rho(r) =
\begin{cases}
\rho_{\rm sol}(r), & r \leq r_t, \\[6pt]
\rho_{\rm CDM}(r), & r \ge r_t ,\label{eq:density}
\end{cases}
\end{equation}
where $r_t = \alpha R_{\rm sol}$ denotes the transition radius at which the soliton core connects to the outer halo. The transition occurs at a fraction of the soliton radius determined by imposing continuity of the density profile. Physically, the final configuration arises from a redistribution of the mass initially contained within a fraction of $R_{\rm sol}$ from a CDM-like power-law profile, eventually reaching virial equilibrium.

The soliton core is described by the density profile \cite{Goodman:2000tg,Boehmer:2007um}
\begin{equation}
\rho_{\rm sol}(r) =
\rho_{c}\,
\frac{\sin(\pi r/R_{\rm sol})}{\pi r/R_{\rm sol}},
\end{equation}
where $\rho_{c}$ is the central density and $R_{\rm sol} = \pi r_{a}$ is the soliton radius. The characteristic scale $r_a$ is given by
\begin{equation}
r_a = \frac{3\lambda}{16\pi G m^{4}},
\end{equation}
where $m$ is the scalar field mass and $\lambda$ is the quartic self-interaction coupling. The density decreases rapidly as $r \to R_{\rm sol}$. This density follows the solution for hydrostatic equilibrium in spherical symmetry, $P\propto \rho^{1+1/n}$, which is the well known analytical solution of the Lane-Emden equation for an $(n = 1)$-polytrope.

The mass enclosed within $R_{\rm sol}$ can be computed from the density profile, yielding
\begin{equation}
M_{\rm sol}= \int_{0}^{R_{\rm sol}} 4\pi r^{2} \rho_{c}
\frac{\sin(\pi r/R_{\rm sol})}{\pi r/R_{\rm sol}}\,dr
= \frac{4}{\pi}\,\rho_{c} R_{\rm sol}^{3}.
\end{equation}

Outside the soliton region, the halo transitions to a CDM-like envelope whose density profile can be approximately described by a power law with logarithmic slope $-12/7$, which differs slightly from the NFW profile\footnote{CDM-like envelopes are often modeled using the NFW profile. Adopting such a profile, however, would introduce additional free parameters that unnecessarily complicate the analysis. The power-law approximation provides a simple fit that captures the main physical behavior of the envelope while preserving a direct connection between the envelope scale and the properties of the soliton core.} $\rho \propto r^{-2}$. The corresponding density profile is
\begin{equation}
\rho_{\rm CDM}(r)= \rho_{0}
\left(\frac{r}{R_{\rm sol}}\right)^{-12/7},
\end{equation}
where $\rho_{0}$ denotes the density of the envelope evaluated at $r = R_{\rm sol}$. This quantity is related to the core density $\rho_{c}$ through two matching conditions. At the transition radius $r_t$, the density profile $\rho(r)$ must be continuous and the enclosed mass must match that of the outer halo, which can be interpreted as conservation of the halo mass during the reconfiguration stage. Solving these constraints simultaneously leads to $\alpha = 0.825$ and $\rho_{0}=0.145 \rho_{c}$. This leaves us with only two independent parameters, $\rho_{c}$ and $R_{\rm sol}$, which characterize the soliton structure and are ultimately determined by the microphysical parameters $m$ and $\lambda$. Heuristically, the core-envelope halo structure follows the  approximate relation
$ \frac{M_{\rm sol}}{R_{\rm sol}}\approx \frac{M_{\rm halo}}{R_{\rm halo}}$, which implies $M_{\rm sol} \propto M_{\rm halo}^{12/21}$ \cite{Chavanis:2019faf}.

\subsection{Numerical procedure}

We summarize the numerical procedure adopted in this work, following closely Ref.~\cite{Figueiredo:2023gas}.

\begin{itemize}

\item The density profile $\rho(r)$ given in Eq.~(\ref{eq:density}) is sampled on a discrete radial grid and interpolated using cubic splines, yielding a smooth representation with continuous first and second derivatives. This interpolated profile is used as input for the numerical integration.

\item The mass function $m(r)$ and the lapse function $f(r)$ are obtained by integrating the corresponding radial equations subject to the boundary conditions specified in Eq.~(\ref{eq:IC_mass}) and Eq.~(\ref{eq:IC_metric}), respectively. The integration is carried out from the black hole horizon $r_h$ up to a cutoff radius $r_{\infty} \gg R$, which effectively approximates spatial infinity.

\item Once the background geometry is reconstructed, we compute the radial pressure $P_{r}$, and the relevant geodesic quantities. This step is performed using an independent numerical scheme, which provides better control over the integration and helps mitigate potential numerical instabilities.

\end{itemize}

\begin{figure}[h]
    \centering
    \includegraphics[width=8.3cm]{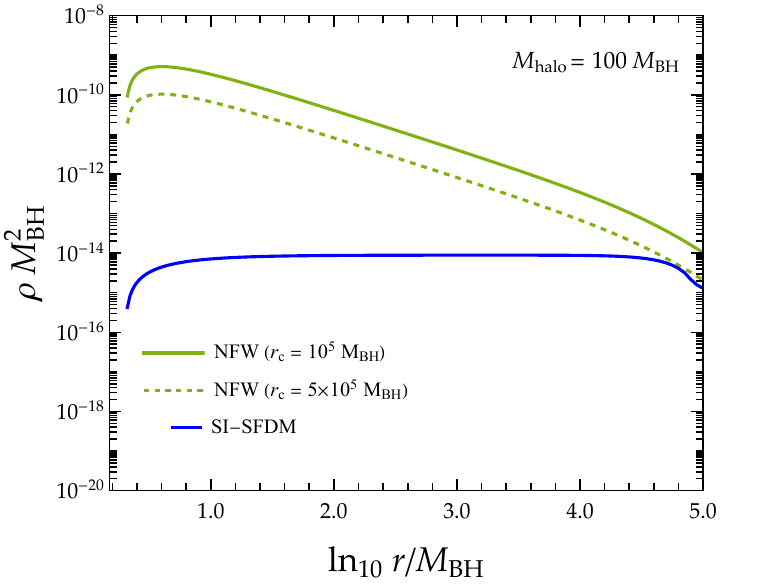}
    \caption{Halo density profiles for $M_{\rm halo} = 100 M_{\rm BH}$, shown for the SI-SFDM and NFW models.}
    \label{fig:rho_0}
\end{figure}

\begin{figure}[h]
    \centering
    \includegraphics[width=8.3cm]{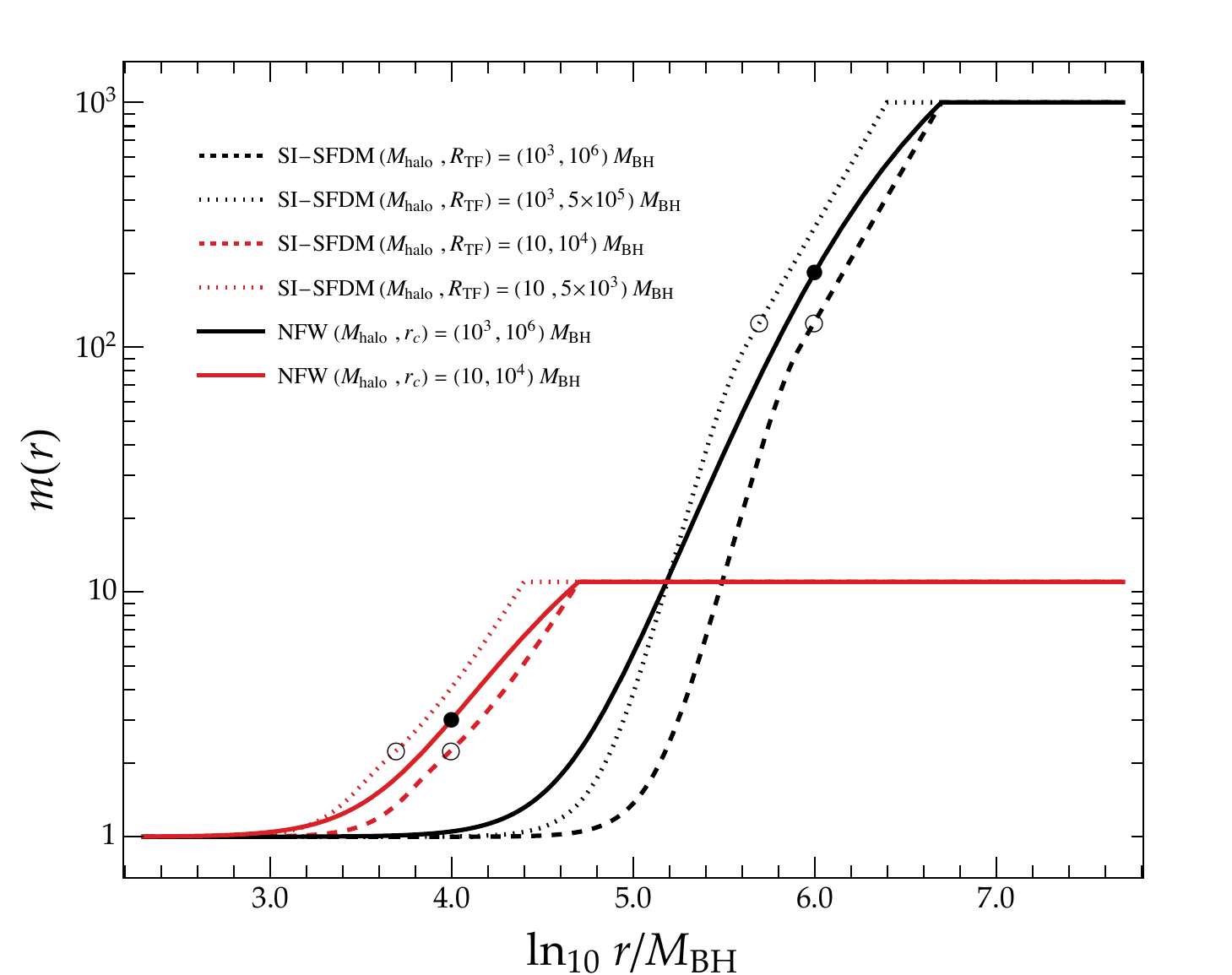} (a) 
    \includegraphics[width=8.3cm]{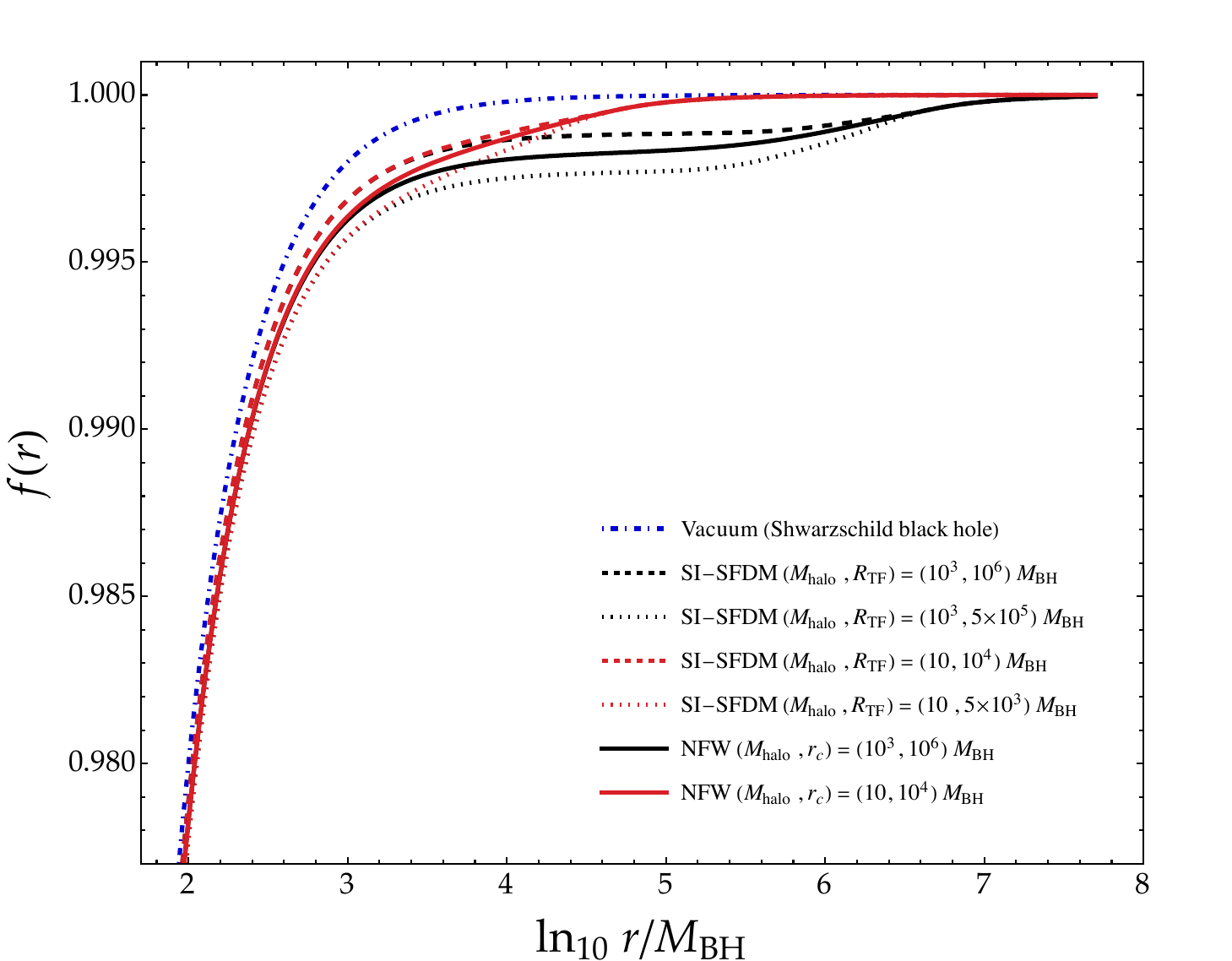} (b)
    \caption{The panels show (a) the profile of $m(r)$ and (b) the corresponding lapse function $f(r)$ for the SI-SFDM and NFW models. The empty circles and black dots in panel (a) indicate, respectively, the values of $R_{\rm TF}$ and $r_c$ for the SI-SFDM and NFW profiles. The blue dot-dashed curve corresponds to a Schwarzschild black hole without any DM halo.}
    \label{fig:m_0f_0}
\end{figure}

In Fig.~\ref{fig:rho_0}, we plot the density profile of the SI-SFDM model and compare it with the NFW profiles for a fixed halo mass $M_{\rm halo}=100M_{\rm BH}$.
Following the same procedure, in Fig.~\ref{fig:m_0f_0} we show the corresponding metric functions $m(r)$ and $f(r)$ for the SI-SFDM and NFW configurations. In panel (a), the mass functions converge to the corresponding total halo mass, either $10M_{\rm BH}$ or $100M_{\rm BH}$, while changes in the core radius mainly affect the inner part of the profiles.
Panel (b) shows that the vacuum Schwarzschild solution gives the upper envelope of the lapse function profiles. The dark matter halo slightly lowers $f(r)$ with respect to the vacuum case, but the deviation remains small, at the level of $\lesssim0.3\%$.
%
%
%

\section{General formalism for photon motion and light deflection angle}\label{sec:ShDe_gen}

In the presence of a black hole, light emitted from a source may undergo deflection due to the gravitational field of the black hole before reaching an observer. Here, we review the general equations required to analyze the light ray trajectories and the deflection angle. 

\subsection{The equations of motion for light rays}\label{subsec:SheDe_gen_1}

Following the standard Lagrangian approach for studying the motion of test particles in a gravitational field, we begin with the Lagrangian \cite{Chandrasekhar:579245}
\begin{equation}
\mathcal{L} = \frac{1}{2} g_{\mu\nu}\dot{x}^{\mu}\dot{x}^{\nu},
\label{eq:L}
\end{equation}
where the overdot denotes differentiation with respect to the affine parameter of the geodesics, denoted by $\tau$. Using this definition, the components of the canonically conjugate momentum are obtained as
\begin{eqnarray}
&& p_t = f(r)\dot t = E,\label{eq:pt}\\
&& p_r = \frac{\dot r}{g(r)},\label{eq:pr}\\
&& p_\theta = r^2\dot\theta,\label{eq:ptheta}\\
&& p_\phi = r^2\sin^2\theta\,\dot\phi = L,\label{eq:pphi}
\end{eqnarray}
in which $E$ and $L$ represent the conserved energy and angular momentum of the test particle, respectively.

The motion can also be analyzed using the Hamilton--Jacobi formalism. The Hamilton--Jacobi equation reads \cite{Carter:1968}
\begin{equation}
\frac{\partial\mathcal{S}}{\partial\tau}
=
-\frac{1}{2}
g^{\mu\nu}
\frac{\partial\mathcal{S}}{\partial x^\mu}
\frac{\partial\mathcal{S}}{\partial x^\nu},
\label{eq:H-J0}
\end{equation}
where $\mathcal{S}$ denotes the Jacobi action. Substituting the metric ansatz \eqref{eq:metric0} into this equation, one obtains
\begin{multline}
-2\frac{\partial\mathcal{S}}{\partial\tau}
=
-\frac{1}{f(r)}
\left(\frac{\partial\mathcal{S}}{\partial t}\right)^2
+
g(r)
\left(\frac{\partial\mathcal{S}}{\partial r}\right)^2
+
\frac{1}{r^2}
\left(\frac{\partial\mathcal{S}}{\partial\theta}\right)^2
\\
+
\frac{1}{r^2\sin^2\theta}
\left(\frac{\partial\mathcal{S}}{\partial\phi}\right)^2 .
\label{eq:H-J1}
\end{multline}

Using the method of separation of variables, and taking into account the conserved quantities introduced above, the Jacobi action can be written as
\begin{equation}
\mathcal{S}
=
\frac{1}{2}m^2\tau
-
Et
+
L\phi
+
\mathcal{S}_r(r)
+
\mathcal{S}_\theta(\theta),
\label{eq:S}
\end{equation}
where $m$ is the rest mass of the particle. Since in this work we are interested in photon trajectories, we set $m=0$. Substituting Eq.~\eqref{eq:S} into Eq.~\eqref{eq:H-J1} leads to
\begin{multline}
0
=
\frac{E^2}{f(r)}
-
g(r)
\left(\frac{\partial\mathcal{S}_r}{\partial r}\right)^2
-
\frac{1}{r^2}
\left[
\frac{L^2}{\sin^2\theta}
+\Q
-
L^2\cot^2\theta
\right]
\\
-
\frac{1}{r^2}
\left[
\left(\frac{\partial\mathcal{S}_\theta}{\partial\theta}\right)^2
-\Q
+
L^2\cot^2\theta
\right],
\label{eq:H-J2}
\end{multline}
where $\Q$ denotes the Carter constant. From this equation the following two relations are obtained
\begin{eqnarray}
&&
r^4 g(r)^2
\left(\frac{\partial\mathcal{S}_r}{\partial r}\right)^2
=
r^4\frac{g(r)}{f(r)}E^2
-
r^2(L^2+\Q)g(r),
\label{eq:H-J3a}\\
&&
\left(\frac{\partial\mathcal{S}_\theta}{\partial\theta}\right)^2
=
\Q-L^2\cot^2\theta .
\label{eq:H-J3b}
\end{eqnarray}

Using the relations \eqref{eq:pt}--\eqref{eq:pphi}, the equations of motion for null geodesics can be written as
\begin{eqnarray}
&& \dot t = \frac{E}{f(r)},\label{eq:tdot}\\
&& \dot r = \pm\frac{\sqrt{\R(r)}}{r^2},\label{eq:rdot}\\
&& \dot\theta = \pm\frac{\sqrt{\Theta(\theta)}}{r^2},\label{eq:thetadot}\\
&& \dot\phi = \frac{L}{r^2\sin^2\theta},
\label{eq:phidot}
\end{eqnarray}
where the $+\;(-)$ sign corresponds to outgoing (ingoing) trajectories. In these expressions we have defined
\begin{subequations}
\begin{align}
\R(r) &= r^4\frac{g(r)}{f(r)}E^2-r^2(L^2+\Q)g(r),\label{eq:R}\\
\Theta(\theta) &= \Q-L^2\cot^2\theta .
\label{eq:Theta}
\end{align}
\end{subequations}

The radial motion of photons is therefore governed by the function $\R(r)$, which plays a role similar to an effective radial potential. The turning points of the trajectories correspond to the radii $r_t$ satisfying $\R(r_t)=0$, which implies $\dot r=0$. These points determine whether a photon coming from infinity is scattered or captured by the compact object.

For the purposes of this work, and without loss of generality, we restrict the motion to the equatorial plane by setting $\theta=\pi/2$, which implies $\Q=0$. In this case the radial function simplifies to
\begin{equation}
\R(r)=r^4\frac{g(r)}{f(r)}E^2-r^2L^2 g(r).
\end{equation}

\begin{figure}[h]
    \centering
    \includegraphics[width=8.3cm]{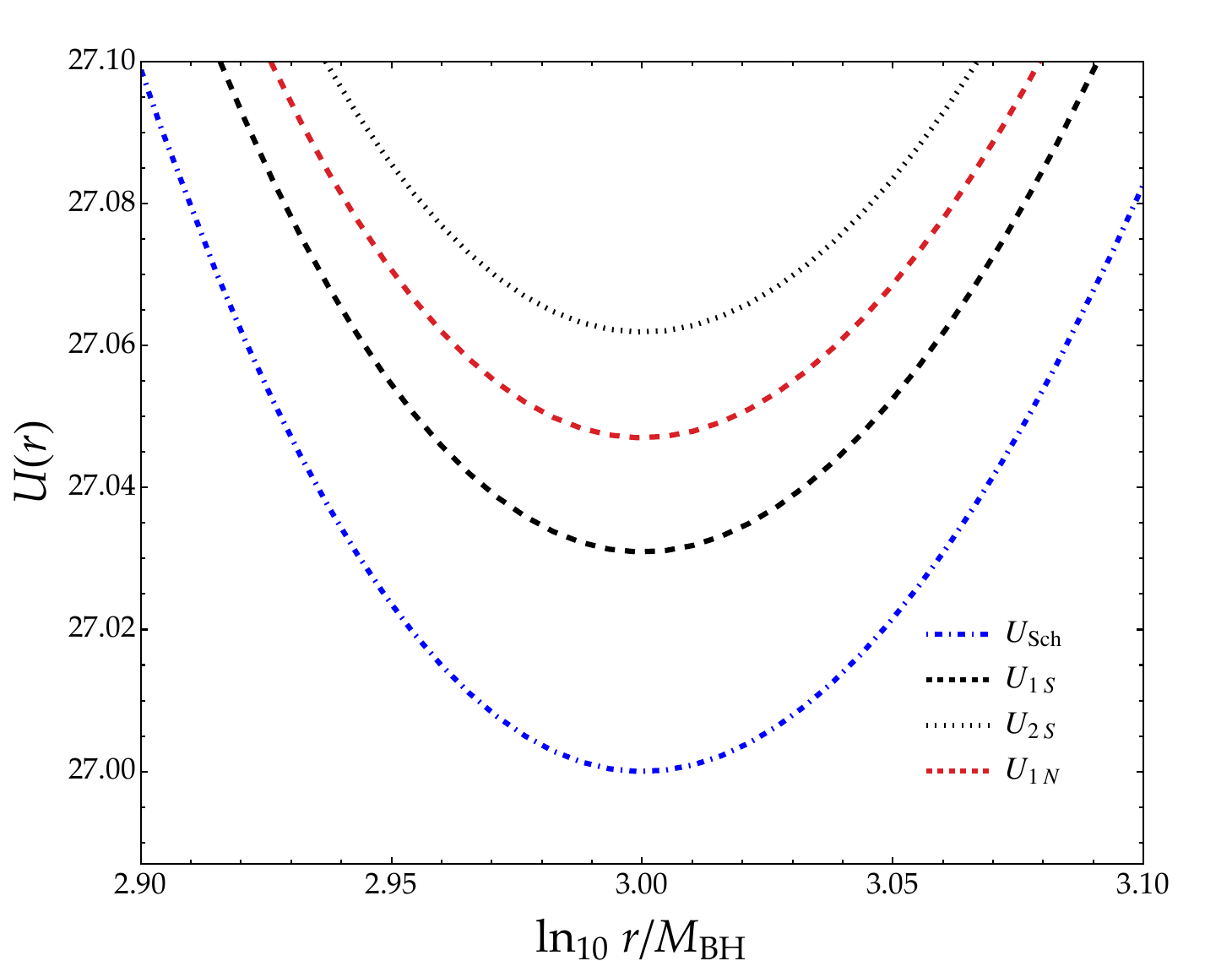} (a) 
     \includegraphics[width=8.3cm]{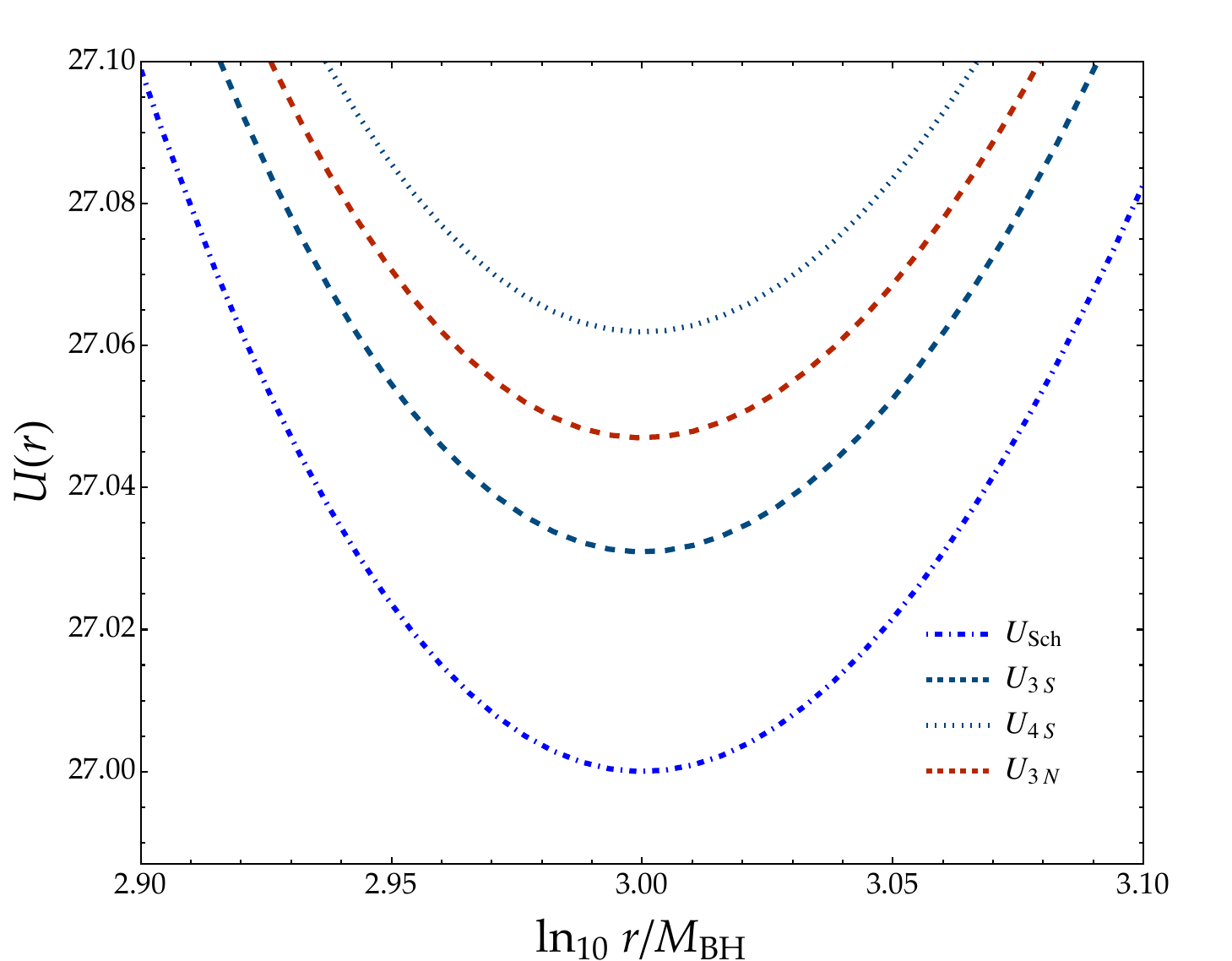} (b) 
    \caption{Radial profile of $U(r)$ for the different halo configurations. The profiles are shown in two panels only to make the comparison clearer. In the displayed region, the Schwarzschild case provides the lower envelope, while the halo configurations slightly increase $U(r)$ due to the dark matter contribution to the mass function. The near-overlaps $U_{1S}\approx U_{3S}$, $U_{2S}\approx U_{4S}$, and $U_{1N}\approx U_{3N}$ are also apparent.}
    \label{fig:U_0}
\end{figure}

It is convenient to introduce the impact parameter $b=L/E$ associated with the photon trajectories. Using this definition, the radial function can be rewritten as
\begin{equation}
\R(r)=E^2 r^2 g(r)\left[U(r)-b^2\right],
\end{equation}
in which $U(r)=r^2/f(r)$. Hence, the physics of photon motion is controlled by the term $U(r)-b^2$. To illustrate this point more clearly, in Fig.~\ref{fig:U_0} we plot the radial profile of $U(r)$ for the different halo models discussed in the previous section. For the present solutions, the Schwarzschild case gives the lowest profile of $U(r)$ in the displayed region. This is consistent with the fact that the dark matter contribution increases the enclosed mass $m(r)$ and slightly lowers $f(r)$ with respect to the vacuum case, as shown in Fig.~\ref{fig:m_0f_0}(b). Since $U(r)=r^2/f(r)$, the halo profiles then lie above the Schwarzschild one. In this sense, the dark matter strengthens the effective gravitational field felt by the photons, although the correction remains very small near the minimum.

It should be noted that, in order to keep the legends of the plots simpler and easier to read, we introduce the notations summarized in Table \ref{tab:1}. 
\begin{table*}[t]
    \centering
    \begin{tabular}{c||c|c}
       notations for metric functions $m(r)$ and $f(r)$ & halo model & adopted values $(M_{\rm halo}, R_{\rm TF}\; \text{or} \; r_c)/M_{\rm BH}$\\
         \hline
       $m_{1S},\, f_{1S}$  & SI-SFDM & $(10^3, 10^6)$ \\
        $m_{2S},\, f_{2S}$  & SI-SFDM & $(10^3, 5\times 10^5)$\\
        $m_{3S},\, f_{3S}$  & SI-SFDM & $(10, 10^4)$ \\
        $m_{4S},\, f_{4S}$  & SI-SFDM & $(10, 5\times10^3)$ \\
         $m_{1N},\, f_{1N}$  & NFW & $(10^3, 10^6)$ \\
         $m_{3N},\, f_{3N}$  & NFW & $(10, 10^4)$ \\
         %
         %
    \end{tabular}
    \caption{The notations used for the metric functions in the forthcoming numerical demonstrations.}
    \label{tab:1}
\end{table*}
Then, the notation for $U(r)$ in Fig. \ref{fig:U_0} follows the same convention as given in Table \ref{tab:1}.
As can be inferred from the diagrams, the profiles possess a minimum, which corresponds to the equation
\begin{equation}
r_p f'(r_p) - 2 f(r_p) = 0,
    \label{eq:Ur_min_eq}
\end{equation}
where $r_p$ corresponds to the radius of unstable photon orbits, and in static spacetimes, identifies the radius of the photon sphere. Furthermore, the impact parameter of photons on such orbits, is the critical impact parameter $b_c$, which is determined as
\begin{equation}
b_c = \frac{r_p}{\sqrt{f(r_p)}},
    \label{eq:bc_eq}
\end{equation}
To quantify the impact of the dark matter halo on photon trajectories, we compute the photon sphere radius $r_p$ and the corresponding critical impact parameter $b_c$ for the different halo models considered in this work. The results, along with their deviations from the Schwarzschild values, are summarized in Table~\ref{tab:rpbc}.
\begin{table}[h]
\centering
\begin{tabular}{c||c|c}
\hline
model & $r_p/M_{\rm BH}$ & $b_c/M_{\rm BH}$ \\
\hline\hline
SBH & $3$ & $3\sqrt{3}$ \\
\hline
$1S$ & $3 + 4.476\times10^{-15}$ & $3\sqrt{3} + 2.975\times10^{-3}$ \\
$2S$ & $3 + 3.581\times10^{-14}$ & $3\sqrt{3} + 5.953\times10^{-3}$ \\
$3S$ & $3 + 4.476\times10^{-11}$ & $3\sqrt{3} + 2.975\times10^{-3}$ \\
$4S$ & $3 + 3.582\times10^{-10}$ & $3\sqrt{3} + 5.954\times10^{-3}$ \\
$1N$ & $3 + 1.565\times10^{-9}$  & $3\sqrt{3} + 4.5213\times10^{-3}$ \\
$3N$ & $3 + 1.565\times10^{-7}$  & $3\sqrt{3} + 4.5208\times10^{-3}$ \\
\hline
\end{tabular}
\caption{
Photon sphere radius $r_p$ and critical impact parameter $b_c$ for different halo models, expressed as deviations from the Schwarzschild black hole (SBH) values $r_p = 3\,M_{\rm BH}$ and $b_c = 3\sqrt{3}\,M_{\rm BH}$. The results show that while the photon sphere radius remains practically unchanged, the critical impact parameter exhibits comparatively larger deviations.
}
\label{tab:rpbc}
\end{table}
As shown in the table, the deviations in the photon sphere radius are extremely small, appearing only at very high numerical precision. This indicates that the near-horizon geometry, which determines $r_p$, remains essentially indistinguishable from that of the Schwarzschild spacetime for all halo models considered. 
In contrast, the critical impact parameter exhibits comparatively larger deviations. This suggests that while the location of the photon sphere is largely insensitive to the presence of the halo, observable quantities related to light propagation, such as the black hole shadow size and strong lensing features, can still encode measurable imprints of the surrounding dark matter distribution. 

\begin{figure}[t]
    \centering
    \includegraphics[width=8.4cm]{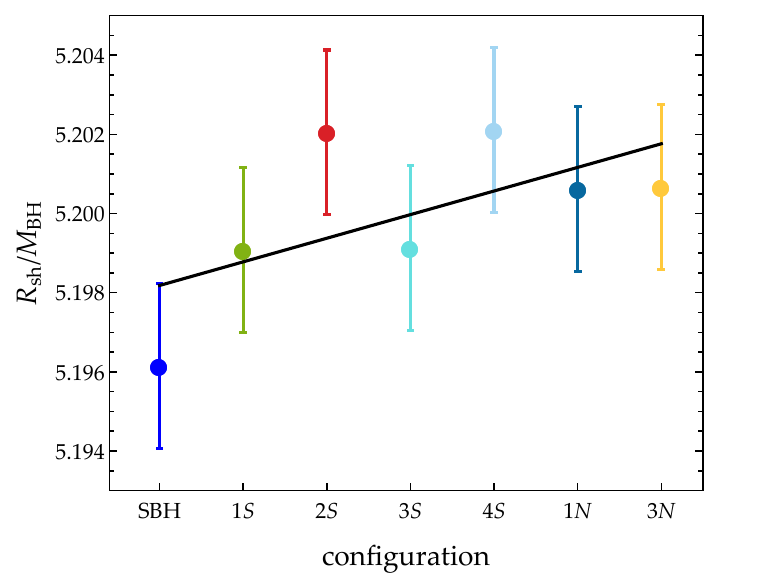} (a)
    \includegraphics[width=8.4cm]{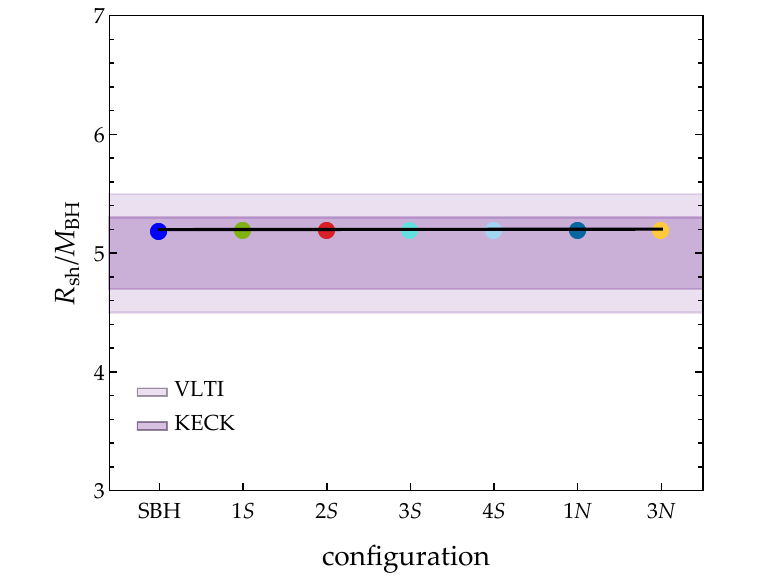} (b)
    \caption{(a) Shadow radius for the different halo configurations considered in this work. The error bars represent the standard deviation of the sample, $\sigma_{R_{\rm sh}} \approx 0.0021$, and therefore indicate the dispersion among the models rather than individual uncertainties. (b) The same shadow radii compared with the observational bounds from the EHT for Sgr A*, where the shaded regions correspond to the VLTI and Keck constraints.}
    \label{fig:Rsh_0}
\end{figure}

The angular radius of the black hole shadow for an observer located at $r_O$ is defined as
\begin{equation}
\sin\theta_{\rm sh} = \frac{b_c \sqrt{f(r_O)}}{r_O},
\label{eq:sin_0}
\end{equation}
which can be equivalently expressed as
\begin{equation}
\sin\theta_{\rm sh} =
\frac{r_p}{r_O}
\sqrt{\frac{f(r_O)}{f(r_p)}}.
\label{eq:sin_1}
\end{equation}
Accordingly, the apparent shadow radius takes the form
\begin{equation}
R_{\rm sh} = r_O \sin\theta_{\rm sh}
= r_p \sqrt{\frac{f(r_O)}{f(r_p)}}.
\label{eq:Rsh_0}
\end{equation}
In the limit $r_O \to \infty$, one recovers $R_{\rm sh} = b_c$. For this case, Fig.~\ref{fig:Rsh_0} shows the shadow radius for the configurations listed in Table~\ref{tab:rpbc}, illustrating the effect of the halo on the observed size.
A linear fit to the data gives $R_{\rm sh}/M_{\rm BH} \approx 5.19759 + 5.97 \times 10^{-4} x$, showing that the variation of the shadow radius across the different halo configurations is of the order $\mathcal{O}(10^{-3})$. This indicates that $R_{\rm sh}$ is nearly constant within the considered set of models and remains very close to the Schwarzschild value throughout the explored configurations.

Moreover, the EHT collaboration has imaged the central black hole in the elliptical galaxy M87 \cite{Akiyama:2019,the_event_horizon_telescope_collaboration_first_2019}, with results that are in excellent agreement with the predictions of general relativity for a Kerr black hole. This was followed by the imaging of Sgr A* \cite{Akiyama:2022,event_horizon_telescope_collaboration_first_2022}, where a bright emission ring was also observed, again consistent with a Kerr geometry.

For Sgr A*, the observational bounds on the shadow radius are given by $4.5\,M_{\rm BH} \lesssim R_{\rm sh} \lesssim 5.5\,M_{\rm BH}$ from Keck and $4.3\,M_{\rm BH} \lesssim R_{\rm sh} \lesssim 5.3\,M_{\rm BH}$ from the VLTI. As shown in panel (b) of Fig.~\ref{fig:Rsh_0}, all the configurations considered here lie well within these observational ranges. In particular, the spread in the theoretical predictions is significantly smaller than the current observational uncertainties, indicating that, although the variation is small, it is not negligible in principle. However, its magnitude remains well below the current observational uncertainties, and therefore the shadow radius alone cannot distinguish between the different halo configurations with present EHT data.


\section{Strong gravitational lensing by the black hole}\label{sec:lensing}

To investigate the gravitational lensing properties around the black hole, it is necessary to study the behavior of null geodesics in the corresponding spacetime. As it is well known, when light rays propagate closer to the black hole, the deflection becomes stronger and the lensing effects are more significant. In order to perform this analysis, we follow the standard strong lensing approach developed in Refs.~\cite{Bozza:2001,Bozza:2002,bozza_strong_2007} (see also Ref.~\cite{bozza_gravitational_2010}).

In general, a static and spherically symmetric spacetime can be written as
\begin{equation}
    \ed s^2 = -A(r)\,\ed t^2 + B(r)\,\ed r^2 + D(r)\left(\ed\theta^2 + \sin^2\theta\,\ed\phi^2\right).
    \label{eq:metric_0}
\end{equation}
In our case, the spacetime metric is given by
\begin{equation}
    g_{\mu\nu} = \left(-f(r),\,\frac{1}{g(r)},\,r^2,\,r^2 \sin^2\theta\right),
    \label{eq:gmunu}
\end{equation}
By comparing the above expressions, one can identify
\begin{equation}
    A(r)=f(r), \qquad B(r)=\frac{1}{g(r)}, \qquad D(r)=r^2.
\end{equation}
Moreover, due to the finite extension of the dark matter halo, the spacetime is asymptotically flat. In particular, in the limit $r\to\infty$, one finds $A(r)\to 1$, $B(r)\to 1$, and $D(r)/r^2 \to 1$.

We assume that the spacetime described by Eq.~\eqref{eq:metric_0} admits at least one photon sphere, where null geodesics follow circular orbits at a constant radius $r_p$. The condition for such orbits is given by $U'(r_p)=0$~\cite{atkinson_light_1965,claudel_geometry_2001,virbhadra_gravitational_2002,Bozza:2002} (see also Ref.~\cite{perlick_calculating_2022}), where for the general metric \eqref{eq:metric_0} one defines
\begin{equation}
    U(r)=\frac{D(r)}{A(r)}.
    \label{eq:V(r)}
\end{equation}
For asymptotically flat spacetimes, one has $U(r)\to\infty$ as $r\to\infty$, and the physically relevant photon sphere corresponds to an extremum at $r=r_p$. As discussed previously, this extremum determines the boundary between captured and scattered null geodesics.

As light rays approach $r_p$, the deflection angle grows rapidly, and photons can wind around the black hole multiple times before reaching the observer. This results in an infinite sequence of images associated with a single source. Following Refs.~\cite{bisnovatyi-kogan_analytical_2022,tsupko_shape_2022,aratore_constraining_2024}, these images are labeled by an integer index $n=0,1,2,3,\dots$, referred to as the order of the image, which counts the number of half-orbits around the black hole.

For a spherically symmetric spacetime, the motion of light can be restricted to the equatorial plane without loss of generality, described by the coordinates $(r,\phi)$. The total change in the azimuthal angle for a photon emitted at $r_S$ and observed at $r_O$ is given by~\cite{bozza_strong_2007}
\begin{equation}
    \Delta\phi=\int_{R_0}^{r_S} b\sqrt{\frac{B(r)}{D(r)\,\mathcal{R}(r,b)}}\,\ed r + \int_{R_0}^{r_O} b\sqrt{\frac{B(r)}{D(r)\,\mathcal{R}(r,b)}}\,\ed r,
    \label{eq:Deltaphi_0}
\end{equation}
where we follow the notation of Ref.~\cite{aratore_constraining_2024} and define
\begin{equation}
    \mathcal{R}(r,b)=\frac{D(r)}{A(r)}-b^2.
    \label{eq:mR}
\end{equation}
Here, $R_0$ denotes the distance of closest approach, determined from the condition $\mathcal{R}(R_0,b)=0$, which allows one to express $R_0$ as a function of $b$. As $R_0\to r_p$, the deflection angle diverges, $\Delta\phi\to\infty$, meaning that photons can orbit the black hole an arbitrarily large number of times. This behavior defines the critical impact parameter $b_c$, obtained from $\mathcal{R}(r_p,b_c)=0$, leading to
\begin{equation}
    b_c=\sqrt{\frac{D(r_p)}{A(r_p)}}.
    \label{eq:bc_0}
\end{equation}
Therefore, Eq.~\eqref{eq:Deltaphi_0} is valid for $b>b_c$. Photons with $b=b_c$ asymptotically approach the photon sphere, while those with $b<b_c$ are captured by the black hole. To cover the full range of impact parameters, it is convenient to introduce the parametrization~\cite{bozza_strong_2007,tsupko_shape_2022,aratore_constraining_2024}
\begin{equation}
    b=b_c\left(1+\epsilon\right),
    \label{eq:bepsilon}
\end{equation}
where $-1\leq\epsilon<\infty$. Since strong lensing effects are dominated by trajectories close to the photon sphere, corresponding to $b\approx b_c$ (or $R_0\approx r_p$), the parameter $\epsilon$ is expected to be small. In this limit, one can write
\begin{equation}
    \epsilon=\frac{b-b_c}{b_c}\ll 1.
    \label{eq:bepsilon_1}
\end{equation}
Note that as $R_0$ approaches $r_p$, the function $\mathcal{R}$ tends to zero, and therefore the integrals in Eq.~\eqref{eq:Deltaphi_0} become divergent. This divergence can be treated using the method introduced in Refs.~\cite{Bozza:2002,bozza_strong_2007}, where a logarithmic approximation is employed near the photon sphere. Within this approach, and assuming that both the source and the observer are located at finite distances, the total azimuthal shift can be written as~\cite{aratore_constraining_2024}
\begin{equation}
\Delta\phi = -\bar{a} \ln\left(\frac{\epsilon}{\eta_S\eta_O}\right)+\bar{\xi}+\pi,
    \label{eq:Deltaphi_1}
\end{equation}
where we have introduced the variable $\eta = 1 - r_p/r$.

The quantity $\bar{\xi}$ is defined as
\begin{equation}
\bar{\xi}=\bar{a}\ln\left(\frac{2\beta_c}{b_c^2}\right)+k_S+k_O-\pi,
    \label{eq:barxi_0}
\end{equation}
with the following definitions:
\begin{subequations}
    \begin{align}
        & \eta_i=1-\frac{r_p}{r_i},\quad i=\{S,O\},\\
        & \bar{a} = r_p\sqrt{\frac{B(r_p)}{A(r_p)\,\beta_c}},\\
        & \beta_c = \frac{r_p^2\Bigl[D''(r_p)A(r_p)-A''(r_p)D(r_p)\Bigr]}{2A^2(r_p)},\\
        & k_i=\int_0^{\eta_i} \mathcal{G}(\eta)\,\mathrm{d}\eta,\label{eq:ki_0}\\
        & \mathcal{G}(\eta)=b_c\sqrt{\frac{B(\eta)}{D(\eta)}}\,\frac{1}{\sqrt{\mathcal{R}(\eta,b_c)}}\frac{r_p}{(1-\eta)^2}
        -\frac{b_c}{\sqrt{\beta_c}}\sqrt{\frac{B(r_p)}{D(r_p)}}\,\frac{r_p}{|\eta|},
    \end{align}
\end{subequations}
where $A(\eta)$, $B(\eta)$, and $D(\eta)$ are obtained from the substitution $r = r_p/(1 - \eta)$ into the metric components in Eq. \eqref{eq:gmunu}. It can be verified that for the vacuum SBH, one gets
\begin{equation}
\mathcal{G}_{\rm SBH}(\eta) = \frac{\sqrt{3}-\sqrt{3-2\eta}}{\eta\sqrt{3-2\eta}},
    \label{eq:G1_SBH}
\end{equation}
as reported in Ref. \cite{Bozza:2002}, which yields $k_i^{\rm SBH} = -2\ln\left(3+\sqrt{3+18 M_{\rm BH}/r_i}\right)$. For the SBH, one has $\eta_i = 1 - 3M_{\rm BH}/r_i$. In this case, the deflection angle can be written as
\begin{equation}
\Delta\phi_{\mathrm{SBH}}-\pi \equiv \alpha_{\mathrm{SBH}} = -\ln\frac{\epsilon}{\eta_S \eta_O}+\xi_{\mathrm{SBH}},
    \label{eq:Deltaphi_SBH}
\end{equation}
where
\begin{subequations}
    \begin{align}
        \xi_{\mathrm{SBH}} = -\pi + 5\ln(6) + k_S^{\mathrm{SBH}} + k_O^{\mathrm{SBH}}.
    \end{align}
\end{subequations}
This result is in agreement with Ref. \cite{bozza_strong_2007}. In the asymptotic limit $r_i \to \infty$, the above expression simplifies to the familiar form
$\alpha_{\mathrm{SBH}} = -\ln\epsilon + \ln\left(216[2-\sqrt{3}]^2\right) - \pi$,
which corresponds to the standard strong-deflection formula for gravitational lensing around an SBH obtained in Refs. \cite{Darwin_gravity_1959,Bozza:2002}. Therefore, one finds $\xi_{\mathrm{SBH}} = -0.4002$, in accordance with the value reported in Ref. \cite{Bozza:2002}.

\begin{figure}[t]
    \centering
    \includegraphics[width=8cm]{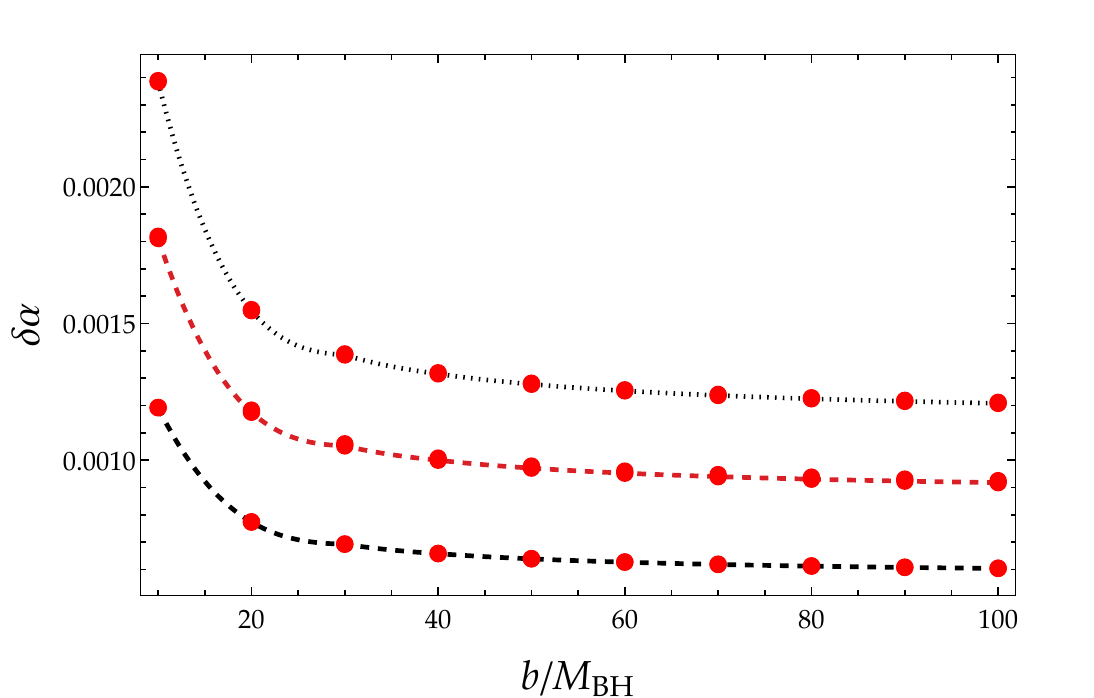} (a)
    \includegraphics[width=8cm]{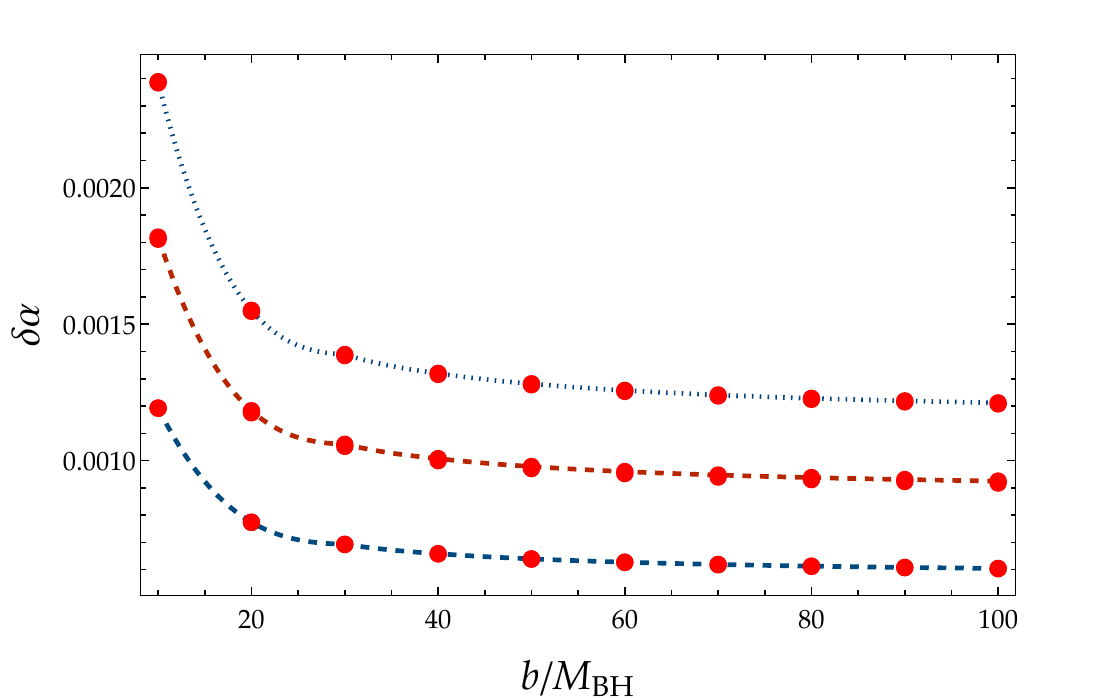} (b)
    \caption{The values for $\delta\alpha$ for different halo configurations. The curve coding follows the same convention as in Fig.~\ref{fig:U_0}.}
    \label{fig:delta_alpha}
\end{figure}

In Fig. \ref{fig:delta_alpha}, we have shown the $b$-profiles of the quantity $\delta\alpha = \alpha-\alpha_{\rm SBH}$, for the different halo profiles under study. 
As expected from the behavior of $U(r)$ in Fig.~\ref{fig:U_0}, the halo contribution slightly increases the effective potential and therefore produces a slightly larger bending angle with respect to the vacuum SBH. The difference is small, but it still separates the set $(1S,2S,1N)$ from $(3S,4S,3N)$.

\section{Lensing observables}\label{sec:lensEq}

In this section, we study the lens equation and the corresponding strong-lensing observables. We then apply these quantities to M87* and Sgr A* in order to estimate the effect of the different halo configurations.

To begin with, we write the general lens equation in the form introduced in \cite{bozza_strong_2007},
\begin{equation}
\phi_O - \phi_S = \Delta\phi \;\mathrm{mod}\; 2\pi,
    \label{eq:lensEq_0}
\end{equation}
where we choose $\phi_O=\pi$ and $\phi_S=0$. Then, by using Eq.~\eqref{eq:Deltaphi_1}, the image positions can be written as
\begin{equation}
\epsilon_n = \eta_O \eta_S \exp\left[\frac{\bar{\xi} - 2n\pi}{\ba}\right],
    \label{eq:epsilon_n}
\end{equation}
where $n$, defined before, shows the number of complete loops made by the light ray around the black hole. Although the exact strong deflection limit corresponds to $n \to \infty$, in practice the case $n=1$ already gives a good approximation. For an observer placed in the asymptotically flat region, the angular separation between the image and the black hole is $\theta=b/r_O$. Then, from Eq.~\eqref{eq:bepsilon}, one gets $\theta=\theta_c(1+\epsilon)$, where $\theta_c=b_c/r_O$ is the angular radius of the black hole shadow. In the regime $r_i \gg r_p$, keeping only the first-order terms in $r_p/r_i$, the total azimuthal shift takes the form \cite{Bozza:2002}
\begin{equation}
\Delta\phi = -\ba \ln\left(\frac{r_O \theta}{b_c} - 1\right) + \bar{\xi}.
    \label{eq:Deltaphi_2}
\end{equation}

Within this approximation, the lens equation becomes \cite{Bozza:2001}
\begin{equation}
\psi = \theta - \frac{r_S}{r_{OS}} \Delta\alpha_n,
    \label{eq:lensEq_1}
\end{equation}
where $r_{OS}=r_O+r_S$ is the distance between the observer and the source, $\Delta\alpha_n=\alpha(\theta)-2n\pi$ is the remaining part of the deflection angle after subtracting the full loops done by the photons, and $\psi$ is the angular position of the source with respect to the black hole. By solving Eq.~\eqref{eq:Deltaphi_2} under the condition $\alpha(\theta_n^0)=2n\pi$, we obtain
\begin{equation}
\theta_n^0 = \frac{b_c}{r_O} \left(1 + \epsilon_n\right),
    \label{eq:thetan0}
\end{equation}
with $\epsilon_n$ given by Eq.~\eqref{eq:epsilon_n}. Expanding $\alpha(\theta)$ around $\theta_n^0$ and introducing $\Delta\theta_n=\theta-\theta_n^0$, the offset angle is found as
\begin{equation}
\Delta\alpha_n = -\frac{\ba r_O}{b_c \epsilon_n} \Delta\theta_n.
    \label{eq:Deltalpha_n}
\end{equation}
Therefore, the lens equation can be recast as
\begin{equation}
\psi = \theta + \left(\frac{\ba r_S r_O}{b_c \epsilon_n r_{OS}}\right) \Delta\theta_n.
    \label{eq:lensEq_2}
\end{equation}
Now, by taking the condition $r_O \gg b_c$, the angular position of the $n$th relativistic image is obtained as \cite{Bozza:2002}
\begin{equation}
\theta_n = \theta_n^0 + \frac{b_c \epsilon_n \left(\psi - \theta_n^0\right) r_{OS}}{\ba\, r_S\, r_O}.
    \label{eq:theta_n}
\end{equation}
This expression clearly shows that the image is exactly aligned with the source when $\psi=\theta_n^0$. Also, the sign of $\psi$ determines on which side of the lens the image is formed: for $\psi>0$, the image appears on the same side, while for $\psi<0$, it appears on the opposite side.

\begin{figure*}[t]
   \centering
   \includegraphics[width=6.cm]{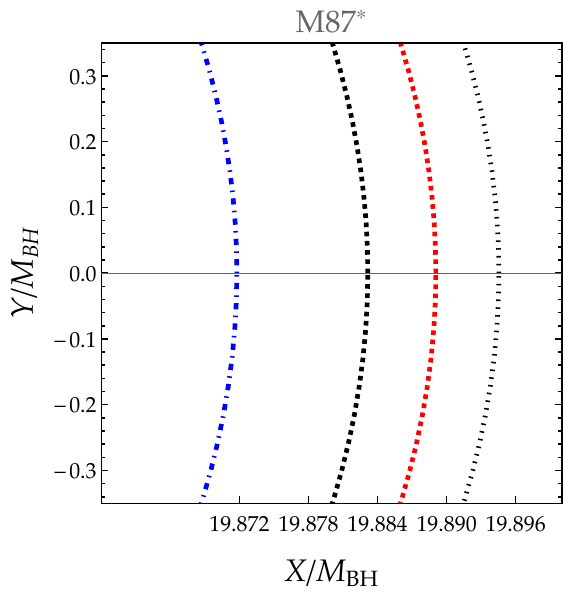} (a)
    \includegraphics[width=6.cm]{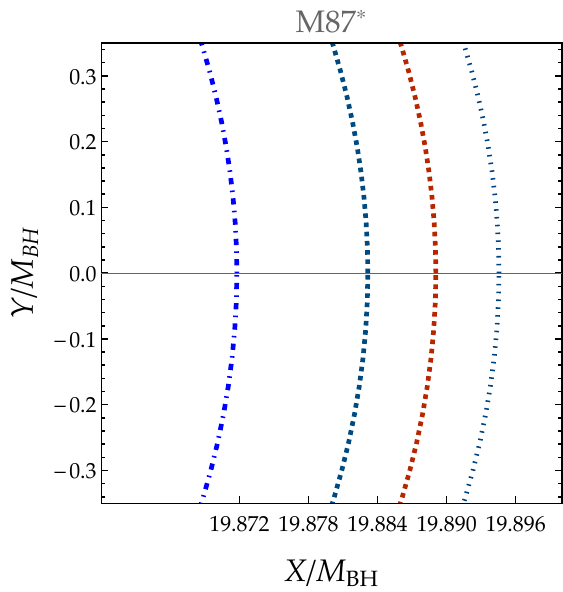} (b)
    \includegraphics[width=6.cm]{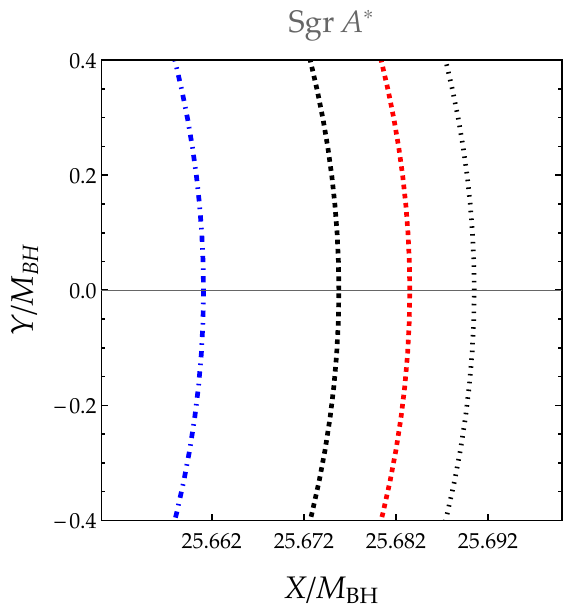} (c)
    \includegraphics[width=6.cm]{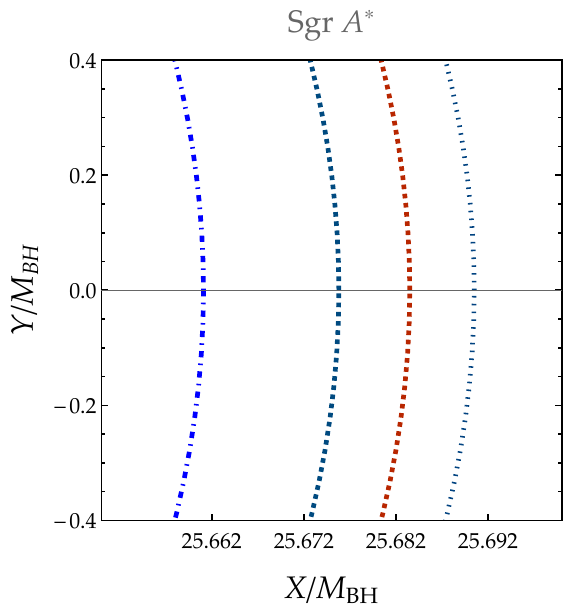} (d)
   \caption{The outermost RERs for M87* (a,b) and Sgr A* (c,d), as black holes with DM halo, for all discussed cases of halo configurations.
   The color coding is the same as that in Fig. \ref{fig:U_0}.}
   \label{fig:Erings}
\end{figure*}

In the special case where the black hole is almost perfectly aligned with the source and the observer, that is $\psi \approx 0$, and when the observer and the lens are equally separated from the source so that $r_{OS}=r_S=2r_O$, the bending of light takes place in all directions and produces the so-called relativistic Einstein rings (RERs) \cite{einstein_lens-like_1936, mellier_probing_1999, bartelmann_weak_2001,petters_relativistic_2003,schmidt_weak_2008, guzik_tests_2010}. In this situation, Eq.~\eqref{eq:theta_n} reduces to \cite{bozza_time_2004}
\begin{equation}
\theta_n^{E} = \left(1 - \frac{b_c \epsilon_n r_{OS}}{\ba\, r_S\, r_O}\right) \theta_n^0.
    \label{eq:theta_nE_0}
\end{equation}
Again, for $r_O \gg b_c$, the angular radius of the $n$th relativistic Einstein ring is simply given by
\begin{equation}
\theta_n^{E} = \frac{b_c \left( 1 + \epsilon_n \right)}{r_O}.
    \label{eq:thetanE_1}
\end{equation}
In order to use these relations in realistic astrophysical systems, we consider the observational data of the supermassive black holes M87* and Sgr A*. In particular, M87*, with mass $(6.5 \pm 0.7)\times10^9\,M_\odot$, is located at a distance $r_O=16.8\,\mathrm{Mpc}$ from Earth \cite{the_event_horizon_telescope_collaboration_first_2019, the_event_horizon_telescope_collaboration_first_2019-1}, whereas Sgr A* has mass $4^{+1.1}_{-0.6}\times10^6\,M_\odot$ and lies at a distance $r_O=7.97\,\mathrm{kpc}$ from Earth \cite{Akiyama:2022, event_horizon_telescope_collaboration_first_2022-1}. The halo information enters these expressions through the numerically reconstructed metric function $f(r)$, and therefore through $r_p$, $b_c$, $\ba$, $\bar{\xi}$, and $\epsilon_n$. Thus, even when the final observational formula is compact, the effect of the halo is already encoded in these strong-lensing quantities. By substituting the observational data into Eq.~\eqref{eq:thetanE_1}, Fig.~\ref{fig:Erings} shows the outermost RERs, corresponding to $n=1$, for both M87* and Sgr A*. These rings are plotted in the celestial coordinates $(X,Y)$ of an observer on Earth and are calculated for the different halo configurations.
The comparison shows that, although small systematic shifts in the RER radius are present, the overall behavior remains unchanged. In both the M87* and Sgr A* cases, the relative variation of the ring radius is of order $\mathcal{O}(10^{-3})$. This estimate refers to the shift of the ring radius itself, not to the relative-magnification quantity $\delta_{r_{\rm mag}}$, whose deviations are much smaller, as shown below.

Another relevant quantity in strong lensing is the magnification of the $n$th relativistic image, defined by \cite{Virbhadra:2000,Bozza:2002}
\begin{equation}
\mu_n = \left.\left(\frac{\psi}{\theta} \frac{\ed \psi}{\ed \theta}\right)^{-1}\right|_{\theta_n^0} = \frac{b_c^2 \epsilon_n \left( 1 + \epsilon_n \right) r_{OS}}{\ba\, \psi\, r_S\, r_O^2}.
    \label{eq:mun}
\end{equation}
This expression shows that the magnification scales as $1/r_O^2$. Therefore, the relativistic images are generally very faint. The outermost image is the brightest one, and the brightness decreases exponentially as the image order increases. Nevertheless, in the limit $\psi \to 0$, which corresponds to an almost perfect alignment between the source and the lens, the images may become strongly magnified. It should also be mentioned that, although the outermost image associated with $\theta_1$ is still separated, the higher-order images accumulate around $\theta_\infty \equiv \theta_n|_{n\to\infty}$ \cite{Bozza:2002}. As already discussed before, this limiting value can be identified as $\theta_\infty=\theta_c$.

The astrophysical study of strong lensing also involves two additional important observables. The first one is the angular separation between the outermost relativistic image and the innermost packed images, which is given by
\begin{equation}
s = \theta_1 - \theta_\infty \approx \theta_\infty \epsilon_1,
    \label{eq:s_0}
\end{equation}
and the second one is the relative magnification of the outermost relativistic image with respect to the collection of all inner relativistic images, which reads \cite{Bozza:2002}
\begin{equation}
r_{\mathrm{mag}} = \frac{\mu_1}{\sum_{n=2}^\infty \mu_n} = 2.5 \log_{10} \left( \exp \left[ \frac{2\pi}{\ba} \right] \right),
    \label{eq:rmag_0}
\end{equation}
and this quantity is independent of the observer distance $r_O$.

With these relations in place, the three observables $\theta_\infty$, $s$, and $r_{\mathrm{mag}}$ can be extracted from astronomical data. The results indicate that the halo configurations considered here induce only very small deviations in these quantities (see Table \ref{tab:shadow_results}).
\begin{table}[t]
\centering
\begin{tabular}{lccccc}
\hline\hline
 & \multicolumn{2}{c}{M87*} & \multicolumn{2}{c}{Sgr A*} &  \\
\cline{2-3} \cline{4-5}
Model & $\theta_{\infty}\,(\mu\mathrm{as})$ & $s\,(\mu\mathrm{as})$ 
      & $\theta_{\infty}\,(\mu\mathrm{as})$ & $s\,(\mu\mathrm{as})$ 
      & $\delta_{r_\mathrm{mag}}$ \\
\hline
SBH  & 19.846900 & 0.0248384 & 25.629100 & 0.0320747 & 0 \\
$1S$    & 19.858300 & 0.0248526 & 25.643700 & 0.0320930 & $-3.29\times10^{-14}$ \\
$2S$   & 19.869700 & 0.0248668 & 25.658400 & 0.0321114 & $-2.74\times10^{-13}$ \\
$3S$   & 19.858300 & 0.0248526 & 25.643700 & 0.0320931 & $-3.44\times10^{-10}$ \\
$4S$   & 19.869700 & 0.0248669 & 25.658400 & 0.0321115 & $-2.75\times10^{-9}$ \\
$1N$    & 19.864200 & 0.0248600 & 25.651400 & 0.0321026 & $-1.07\times10^{-8}$ \\
$3N$   & 19.864200 & 0.0248602 & 25.651400 & 0.0321029 & $-1.07\times10^{-6}$ \\
\hline\hline
\end{tabular}
\caption{Angular radius $\theta_{\infty}$ and separation $s$ for M87* and Sgr A* for different configurations. The parameter $\delta_{r_\mathrm{mag}}$ represents the deviation of the relative magnification with respect to the Schwarzschild case, for which $r_{\rm mag}=6.8219$.}
\label{tab:shadow_results}
\end{table}
The results presented in the table show that the angular radius $\theta_\infty$ and the separation $s$ exhibit only very small variations across the different halo configurations, for both M87* and Sgr A*. The dark matter contribution enters these quantities through the metric function $f(r)$, and therefore through $b_c$, $\ba$, $\bar{\xi}$, and $\epsilon_1$. Thus, $\theta_\infty$, $s$, and $r_{\rm mag}$ are not independent of the halo, although their final expressions are compact. In particular, the changes in $\theta_\infty$ are at the level of $\mathcal{O}(10^{-3})$, while the corresponding variations in $s$ are even smaller, remaining at the level of $\mathcal{O}(10^{-5})$--$\mathcal{O}(10^{-6})$. A similar behavior is reflected in $r_{\mathrm{mag}}$, whose deviation with respect to the Schwarzschild case is extremely small, from $\mathcal{O}(10^{-14})$ up to $\mathcal{O}(10^{-6})$. The different orders in $\delta_{r_{\rm mag}}$ are mainly related to the effective compactness of the adopted halo profile. For example, although $1S$ has a larger total halo mass, it is more extended and its effect near the photon sphere is strongly suppressed. The $4S$ profile is more compact, and therefore gives a larger local correction to the strong-lensing coefficients, even with a smaller total halo mass.

At this stage, it becomes evident that the observables $\theta_\infty$ and $s$ are only weakly affected by the presence of the halo configurations considered in this work. Although small deviations are present, their magnitude remains below the level required for a clear observational distinction.

Motivated by this limitation, we proceed to explore additional strong gravitational lensing observables that are known to exhibit a higher sensitivity to the underlying spacetime geometry. In particular, we focus on quantities related to higher-order images, magnification patterns, and time delays, which provide a more stringent test of the deviations induced by the halo structure.

\section{Beyond the primary strong-lensing observables}\label{sec:higher_order}

Although the observables $\theta_\infty$, $s$, and $r_{\mathrm{mag}}$ already provide a useful first characterization of the strong-lensing behavior, the results obtained above indicate that they remain only weakly sensitive to the halo configurations considered in this work. This motivates us to go one step further and examine additional strong-lensing quantities within the same finite-distance framework. In particular, we now investigate finite-order relativistic image positions, individual image magnifications, time delays, and the gap between higher-order rings, with the aim of identifying observables that may display a more pronounced dependence on the halo structure.

Before presenting the separate observables, we fix the notation used to measure the departure from the vacuum Schwarzschild case. For any quantity $X$, we denote the absolute correction by $\delta X=X-X^{\rm SBH}$, while the corresponding relative correction is written as $\delta X/X^{\rm SBH}$. When the quantity itself is already a difference, such as $\Delta T_{2,1}$, we write the absolute correction as $\delta(\Delta T_{2,1})=\Delta T_{2,1}-\Delta T_{2,1}^{\rm SBH}$.

\subsection{Finite-order relativistic image positions}\label{subsec:finite-order}

While the observable $\theta_\infty$ describes the asymptotic accumulation point of the relativistic images, additional information can be extracted from the finite-order image positions themselves. In the present finite-distance formalism, these quantities are already determined by Eqs.~\eqref{eq:bepsilon}, \eqref{eq:Deltaphi_1}, \eqref{eq:ki_0}, \eqref{eq:bc_0}, \eqref{eq:bepsilon_1}, and, more directly, by Eqs.~\eqref{eq:thetan0} and \eqref{eq:theta_n}, which provide the angular position of the $n$th relativistic image.

The importance of these observables lies in the fact that the first few relativistic images are not yet fully merged into the limiting value $\theta_\infty$. Therefore, unlike $\theta_\infty$, which mainly captures the asymptotic structure of the image sequence, the positions $\theta_n$ with finite $n$ retain more detailed information about the spacetime geometry close to the photon sphere. In particular, since $\theta_n$ depends on the quantities $\bar{a}$, $\bar{\xi}$, $\eta_S$, and $\eta_O$, it is expected to be more sensitive to small differences among the halo configurations.

In what follows, we focus on the first few relativistic images, especially $n=1,2,3$, and compute their angular positions for the different halo models considered in this work. For this finite-order analysis, we use the deviation with respect to the Schwarzschild case,
\begin{equation}
\delta \theta_n = \theta_n - \theta_n^{\rm SBH}.
\end{equation}
This quantity allows us to determine whether the finite-order relativistic images provide a more sensitive probe of the halo structure than the already studied observables $\theta_\infty$ and $s$.

It is also useful to consider the separation between successive relativistic images, defined as
\begin{equation}
\Delta \theta_{n,n+1} = \theta_n - \theta_{n+1},
\end{equation}
which measures how rapidly the image sequence accumulates toward $\theta_\infty$. Since the accumulation pattern is controlled by the strong-lensing coefficients, this quantity may reveal differences between halo configurations even in cases where $\theta_\infty$ itself remains nearly unchanged.

In order to proceed with the calculations, it is worth noting that, while we retain the condition $r_{OS}=r_S=2r_O$, we no longer impose the specific choice $\psi=\theta_n^0$. This is because such a condition enforces the RER configuration and effectively reduces the general expression for the image positions to the corresponding ring formulas. In this regard, Eq. \eqref{eq:theta_n} reduces to
\begin{equation}
\theta_n = \theta_n^0 + \frac{b_c\epsilon_n\left(\psi-\theta_n^0\right)}{\bar{a}\, r_O}.
    \label{eq:theta_n_r}
\end{equation}
Therefore, one has to choose a suitable value for the angle $\psi$: it should be small enough to remain in the strong-lensing regime, but not so small that the configuration effectively returns to the ring limit. The cleanest practical choice is to set $\psi$ as a fixed fraction of the Schwarzschild shadow angle, i.e.,
\begin{equation}
\psi = \kappa \theta_\infty^{\rm SBH},
    \label{eq:psi_0}
\end{equation}
with a small constant $\kappa$. A good first choice is $\kappa = 0.1$. 

Tables~\ref{tab:finite_images_M87} and \ref{tab:finite_images_SgrA} present the finite-order relativistic image positions for the BH+DM configurations associated with M87* and Sgr A*, respectively. 
\begin{table*}[t]
\centering
\begin{tabular}{lccccc}
\hline\hline
Model & $\theta_1\,(\mu{\rm as})$ & $\theta_2\,(\mu{\rm as})$ 
& $\delta\theta_1\,(\mu{\rm as})$ & $\delta\theta_2\,(\mu{\rm as})$ 
& $\Delta\theta_{12}\,(\mu{\rm as})$ \\
\hline
SBH & 19.8718 & 19.8470 & 0 & 0 & 0.0247920 \\
$1S$  & 19.8832 & 19.8583 & 0.0113765 & 0.0113623 & 0.0248062 \\
$2S$  & 19.8945 & 19.8697 & 0.0227652 & 0.0227368 & 0.0248204 \\
$3S$  & 19.8832 & 19.8583 & 0.0113778 & 0.0113636 & 0.0248062 \\
$4S$  & 19.8945 & 19.8697 & 0.0227704 & 0.0227419 & 0.0248205 \\
$1N$  & 19.8891 & 19.8643 & 0.0172910 & 0.0172694 & 0.0248136 \\
$3N$  & 19.8891 & 19.8643 & 0.0172890 & 0.0172672 & 0.0248138 \\
\hline\hline
\end{tabular}
\caption{Finite-order relativistic image positions for M87*. The quantities $\delta\theta_1$ and $\delta\theta_2$ are measured with respect to the Schwarzschild black hole case, while $\Delta\theta_{12}=\theta_1-\theta_2$.}
\label{tab:finite_images_M87}
\end{table*}
\begin{table*}[t]
\centering
\begin{tabular}{lccccc}
\hline\hline
Model & $\theta_1\,(\mu{\rm as})$ & $\theta_2\,(\mu{\rm as})$ 
& $\delta\theta_1\,(\mu{\rm as})$ & $\delta\theta_2\,(\mu{\rm as})$ 
& $\Delta\theta_{12}\,(\mu{\rm as})$ \\
\hline
SBH & 25.6611 & 25.6291 & 0 & 0 & 0.0320148 \\
$1S$  & 25.6758 & 25.6438 & 0.0146908 & 0.0146725 & 0.0320331 \\
$2S$  & 25.6905 & 25.6585 & 0.0293975 & 0.0293608 & 0.0320515 \\
$3S$  & 25.6758 & 25.6438 & 0.0146925 & 0.0146742 & 0.0320331 \\
$4S$  & 25.6905 & 25.6585 & 0.0294042 & 0.0293674 & 0.0320516 \\
$1N$  & 25.6835 & 25.6514 & 0.0223284 & 0.0223006 & 0.0320426 \\
$3N$  & 25.6835 & 25.6514 & 0.0223259 & 0.0222978 & 0.0320429 \\
\hline\hline
\end{tabular}
\caption{Finite-order relativistic image positions for Sgr A*. The quantities $\delta\theta_1$ and $\delta\theta_2$ are measured with respect to the SBH case, while $\Delta\theta_{12}=\theta_1-\theta_2$.}
\label{tab:finite_images_SgrA}
\end{table*}
The results show that the finite-order image positions follow the same general pattern observed for the previous shadow and lensing quantities. The largest deviations from the Schwarzschild case are obtained for the $2S$ and $4S$ configurations, while the $1S$ and $3S$ models remain almost degenerate. This behavior is mainly controlled by the effective compactness of the halo in the region probed by the photon trajectories. The pairs $1S$--$3S$ and $2S$--$4S$ have very similar compactness ratios and therefore generate nearly the same corrections to the strong-lensing quantities. The $2S$ and $4S$ cases are more compact than $1S$ and $3S$, which is why their deviations are larger. A similar near-degeneracy is also found between the $1N$ and $3N$ configurations, whose values of $\theta_1$ and $\theta_2$ coincide at the displayed precision. This indicates that the finite-order image positions reveal small systematic shifts induced by the halo, although they are still not sufficient by themselves to fully distinguish all configurations.

\begin{figure*}[t]
    \centering
    \includegraphics[width=8cm]{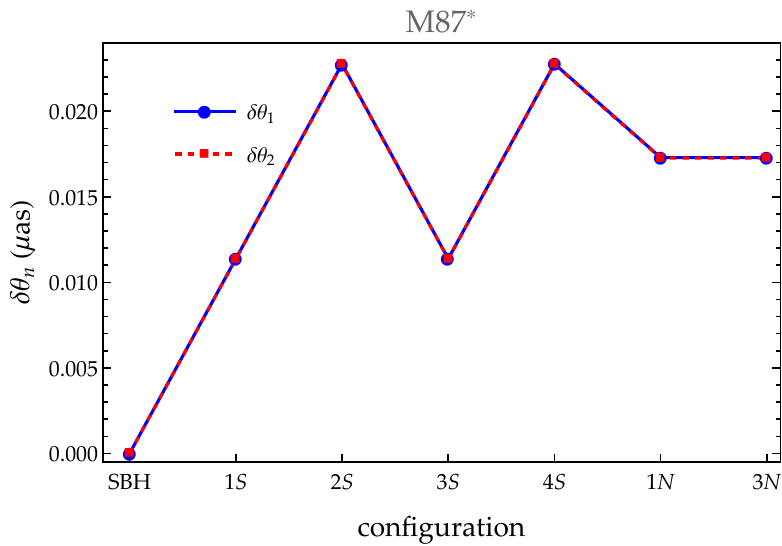} \qquad
    \includegraphics[width=8cm]{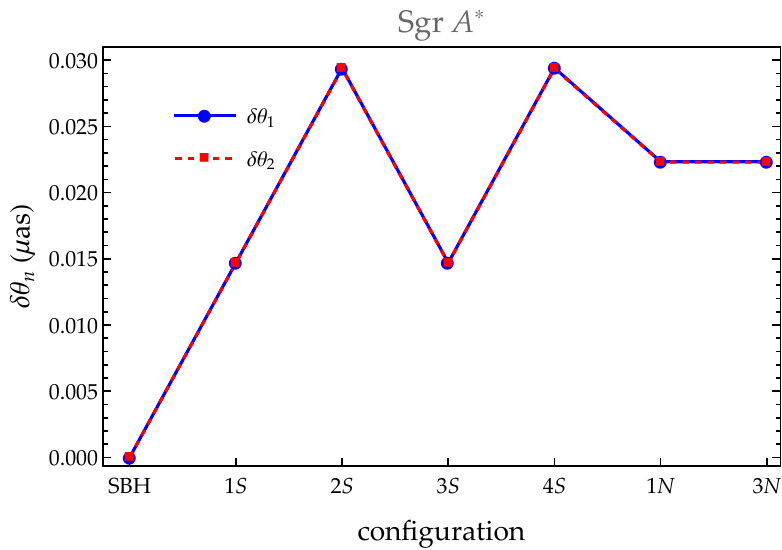} 
    \caption{Deviations of the finite-order relativistic image positions, $\delta\theta_n=\theta_n-\theta_n^{\rm SBH}$, for $n=1$ and $n=2$, shown for different halo configurations. The left and right panels correspond to M87* and Sgr A*, respectively.}
    \label{fig:finite_image_deviations}
\end{figure*}

In order to better quantify the deviations from the Schwarzschild case, we plot the quantities $\delta\theta_n$ for the first two relativistic images, namely $n=1$ and $n=2$, for all halo configurations. The results are shown in Fig.~\ref{fig:finite_image_deviations} for both M87* and Sgr A*.
As can be seen, $\delta\theta_1$ and $\delta\theta_2$ follow almost the same behavior across all configurations. The largest deviations correspond to the $2S$ and $4S$ models, while the pairs $1S$--$3S$ and $1N$--$3N$ remain nearly degenerate. This near overlap indicates that the halo contribution produces mainly a common shift of the first relativistic images, rather than strongly modifying their relative spacing.

\begin{figure*}[t]
    \centering
    \includegraphics[width=8cm]{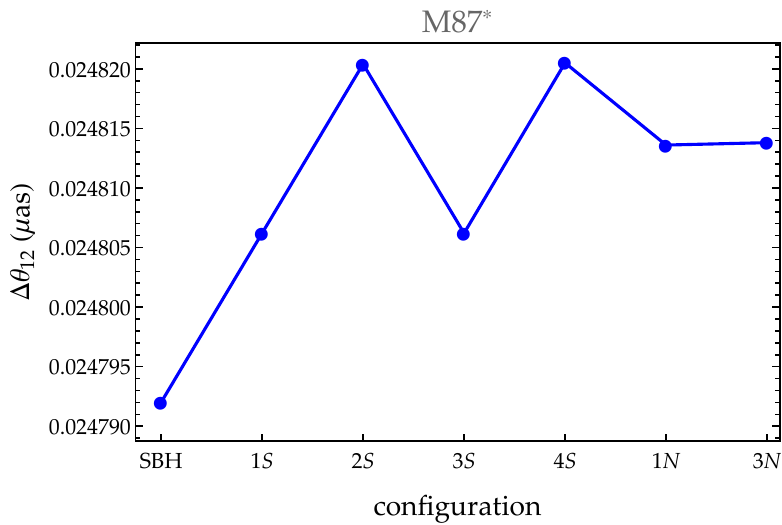} \qquad
    \includegraphics[width=8cm]{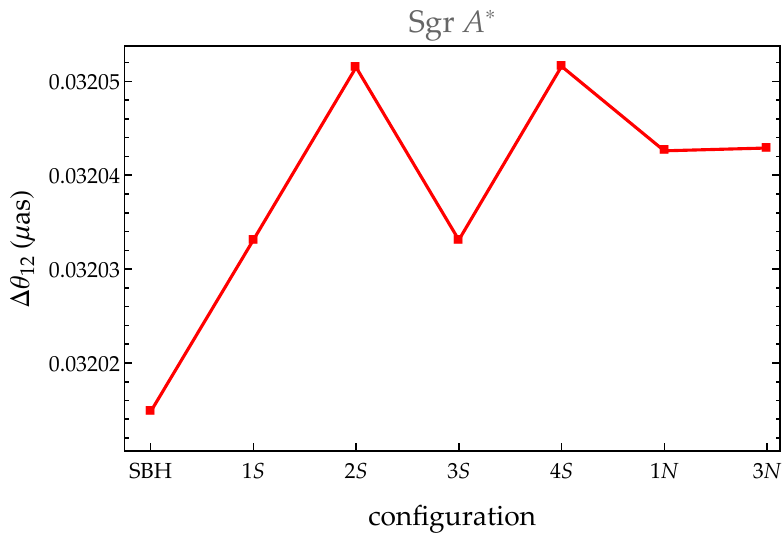} 
    \caption{Separation between the first two finite-order relativistic images, $\Delta\theta_{12}=\theta_1-\theta_2$, for different halo configurations. The left and right panels correspond to M87* and Sgr A*, respectively.}
    \label{fig:Delta_theta12}
\end{figure*}

To examine this point more directly, we also consider the separation between the first two relativistic images, $\Delta\theta_{12}=\theta_1-\theta_2$. The corresponding results are shown in Fig.~\ref{fig:Delta_theta12}.
The behavior of $\Delta\theta_{12}$ follows the same ordering observed in the image-position deviations, with the largest values again obtained for the $2S$ and $4S$ configurations. However, the total variation remains extremely small, at the level of a few $10^{-5}\,\mu{\rm as}$. Therefore, although the separation between the first two relativistic images slightly reduces the common-shift degeneracy seen in Fig.~\ref{fig:finite_image_deviations}, it still does not provide a strong observational discriminator by itself. This motivates us to move to quantities that are expected to be more sensitive to the image hierarchy, such as the magnification pattern and the time delays between relativistic images.

\subsection{Beyond geometric observables: toward stronger discriminators}\label{subsec:beyond_geom}

The results obtained so far indicate that both the finite-order image positions and their separations are only weakly affected by the halo configurations. In particular, the near-degeneracy observed in $\delta\theta_n$ and the extremely small variations in $\Delta\theta_{12}$ suggest that purely geometric observables are not sufficient to clearly distinguish between the considered models. Therefore, it is necessary to move toward quantities that depend more sensitively on the propagation of light in the strong-field region. In this context, observables such as the magnification of relativistic images and the associated time delays provide more promising probes, as they directly depend on the full structure of the null geodesics near the photon sphere.

\subsection{Magnification of finite-order relativistic images}

The results obtained from the finite-order image positions indicate that the halo configurations mainly induce a common shift in the relativistic image sequence. For this reason, we now consider the magnification of these images, since this quantity depends more directly on the strong-lensing coefficients and may therefore provide a more sensitive probe of the halo structure. Following the standard strong-deflection treatment~\cite{Virbhadra:2000,Bozza:2002,Bozza:Scarpetta:2007}, the magnification of the $n$-th relativistic image is given by Eq.~\eqref{eq:mun}.

In the present analysis, we keep the same geometrical setup adopted in the previous subsection, as well as that in Eq. \eqref{eq:psi_0}.
Therefore, the magnification is evaluated as
\begin{equation}
\mu_n = \frac{b_c^2\epsilon_n(1+\epsilon_n)}{\kappa\,\ba\,b_c^{\rm SBH}\,r_O},    \label{eq:mu_n_new}
\end{equation}
for the same source position in all configurations.

Since the absolute value of $\mu_n$ depends on the chosen source position, it is more useful to compare normalized quantities. We therefore introduce the relative magnification deviation
\begin{equation}
    \delta\mu_n =
    \frac{\mu_n-\mu_n^{\rm SBH}}{\mu_n^{\rm SBH}},
\end{equation}
where $\mu_n^{\rm SBH}$ denotes the corresponding Schwarzschild value. In addition, we consider the magnitude difference between the first two relativistic images,
\begin{equation}
    \Delta m_{12}=2.5\log_{10}\left(\frac{\mu_1}{\mu_2}\right).
\end{equation}
These quantities allow us to examine whether the halo configurations modify the brightness hierarchy of the relativistic image sequence more efficiently than the image positions themselves.

\begin{figure}[t]
    \centering
    \includegraphics[width=8.5cm]{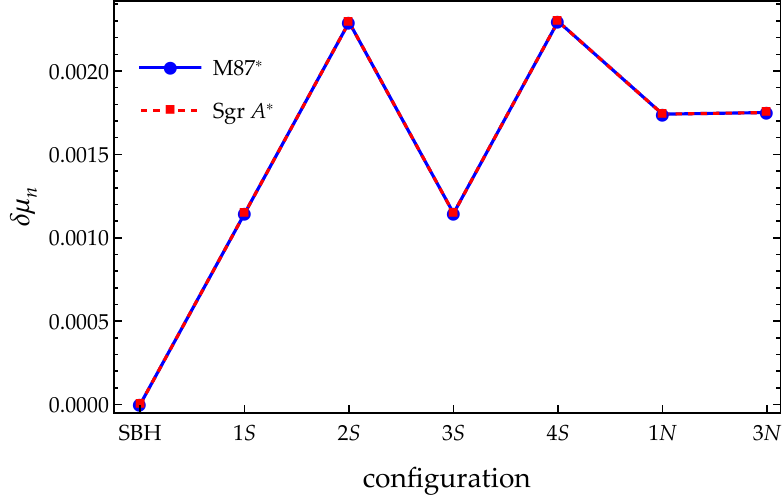}
    \caption{Relative magnification deviation $\delta\mu_n$ for the first two relativistic images, $n=1,2$, for different halo configurations. The values for M87* and Sgr A* are practically indistinguishable, differing only at the level of $\mathcal{O}(10^{-14})$.}
    \label{fig:delta_mu_n}
\end{figure}

Using Eq.~\eqref{eq:mu_n_new}, we calculate the magnification of the first two relativistic images, $n=1,2$, for all halo configurations. The corresponding relative deviations are shown in Fig.~\ref{fig:delta_mu_n}.
As can be seen, the relative magnification deviations follow the same hierarchy already observed in the image-position analysis. The largest deviations are obtained for the $2S$ and $4S$ configurations, while the pairs $1S$--$3S$ and $1N$--$3N$ remain nearly degenerate. Moreover, for each configuration, the values of $\delta\mu_1$ and $\delta\mu_2$ coincide within numerical precision. This means that the halo modifies the overall magnification scale of the first relativistic images, but does not significantly alter their relative brightness hierarchy.

In addition, we find that $\Delta m_{12}$ remains constant for all configurations, with $\Delta m_{12}\simeq 6.82324$. This happens because the first two magnifications are rescaled almost by the same factor in each halo model, so their ratio is nearly unchanged. The logarithm then makes this already small difference even less visible. Therefore, although the magnification contains systematic halo-induced corrections at the level of $\mathcal{O}(10^{-3})$, it does not provide an efficient discriminator among the halo models in the present setup. This further motivates the study of time-delay observables, which are expected to be more sensitive to the accumulated photon propagation in the strong-field region.

\subsection{Time delay between relativistic images}

As shown in the previous subsections, the finite-order image positions, their separations, and the magnification pattern exhibit only small deviations among the considered halo configurations. We therefore consider the time delay between relativistic images as an additional strong-lensing observable. This quantity is physically important because photons forming different relativistic images wind around the black hole a different number of times before reaching the observer, and therefore they accumulate different travel times~\cite{Bozza:2004,Bozza:Scarpetta:2007,Wang:2025}.

For a static and spherically symmetric spacetime, the travel time of a photon whose closest approach is $R$ can be written in terms of the null geodesic equations. In the strong-deflection regime, the dominant contribution to the time delay between two relativistic images comes from the extra loops performed near the photon sphere. Accordingly, the leading time delay between two images of orders $n$ and $m$ is given by~\cite{Bozza:2004,Wang:2025}
\begin{equation}
    \Delta T_{n,m}
    \simeq 2\pi |n-m|\,b_c ,
    \label{eq:time_delay_nm}
\end{equation}
where $b_c$ is the critical impact parameter associated with the photon sphere. In particular, for two consecutive relativistic images, one obtains
\begin{equation}
    \Delta T_{2,1}
    \simeq 2\pi b_c = 2\pi \theta_\infty r_O.
    \label{eq:time_delay_21_geom}
\end{equation}
This expression is written in geometrized units, where the black hole mass sets the unit of time. Therefore, if $b_c$ is expressed in units of $M_{\rm BH}$, the physical time delay is obtained as
\begin{equation}
    \Delta T_{2,1}^{\rm phys}
    =
    2\pi b_c\,\frac{G M_{\rm BH}}{c^3}.
    \label{eq:time_delay_21_phys}
\end{equation}
Equivalently, in minutes,
\begin{equation}
    \Delta T_{2,1}^{\rm min}
    =
    \frac{2\pi b_c}{60}\,
    \frac{G M_{\rm BH}}{c^3}.
    \label{eq:time_delay_21_min}
\end{equation}

For the numerical comparison, we distinguish between the absolute correction and the relative deviation with respect to the Schwarzschild case. Since the time-delay table uses both quantities, we define them explicitly here. The absolute halo-induced correction is
\begin{equation}
    \delta(\Delta T_{2,1})
    =
    \Delta T_{2,1}-\Delta T_{2,1}^{\rm SBH},
    \label{eq:absolute_time_delay_deviation}
\end{equation}
while the corresponding relative deviation is defined as
\begin{equation}
    \delta T_{2,1}
    =
    \frac{\Delta T_{2,1}-\Delta T_{2,1}^{\rm SBH}}
    {\Delta T_{2,1}^{\rm SBH}}.
    \label{eq:delta_time_delay}
\end{equation}
Since the leading time delay is directly proportional to $b_c$, the relative deviation can also be written as
\begin{equation}
    \delta T_{2,1}
    =
    \frac{b_c-b_c^{\rm SBH}}{b_c^{\rm SBH}}.
    \label{eq:delta_time_delay_bc}
\end{equation}
Thus, at leading order, $\delta T_{2,1}$ measures the fractional change of the critical impact parameter, while $\delta(\Delta T_{2,1})$ gives the corresponding correction in physical units.

Using Eqs.~\eqref{eq:time_delay_21_phys}--\eqref{eq:delta_time_delay_bc}, we compute the time delay between the first two relativistic images for all halo configurations. In continuity with the previous subsections, we consider both M87* and Sgr A*. The resulting values of $\Delta T_{2,1}$, together with their deviations from the Schwarzschild case, are summarized in Table~\ref{tab:time_delay}.
\begin{table*}[t]
\centering
\begin{tabular}{lcccccc}
\hline\hline
Model & $b_c$ & $\Delta T_{2,1}^{\rm M87*}$ & $\delta(\Delta T_{2,1})^{\rm M87*}$ 
& $\Delta T_{2,1}^{\rm Sgr A*}$ & $\delta(\Delta T_{2,1})^{\rm Sgr A*}$ 
& $\delta T_{2,1}$ \\
 &  & (min) & (min) & (min) & (min) &  \\
\hline
SBH & 5.196152 & 17421.54 & 0 & 11.52502 & 0 & 0 \\
$1S$ & 5.199127 & 17431.51 & 9.9737 & 11.53162 & 0.006598 & $5.7249\times10^{-4}$ \\
$2S$ & 5.202105 & 17441.50 & 19.9582 & 11.53822 & 0.013203 & $1.1456\times10^{-3}$ \\
$3S$ & 5.199128 & 17431.51 & 9.9748 & 11.53162 & 0.006599 & $5.7256\times10^{-4}$ \\
$4S$ & 5.202106 & 17441.50 & 19.9627 & 11.53822 & 0.013206 & $1.1459\times10^{-3}$ \\
$1N$ & 5.200674 & 17436.70 & 15.1589 & 11.53505 & 0.010028 & $8.7013\times10^{-4}$ \\
$3N$ & 5.200673 & 17436.70 & 15.1571 & 11.53505 & 0.010027 & $8.7002\times10^{-4}$ \\
\hline\hline
\end{tabular}
\caption{Time delay between the first two relativistic images for different halo configurations. The quantities $\delta(\Delta T_{2,1})$ denote the absolute corrections with respect to the Schwarzschild case, while $\delta T_{2,1}$ is the corresponding relative deviation.}
\label{tab:time_delay}
\end{table*}
As shown in Table~\ref{tab:time_delay}, the relative deviation $\delta T_{2,1}$ remains at the level of $\mathcal{O}(10^{-3})$ for all configurations, indicating that the halo induces only a small fractional correction to the time delay. However, due to the mass scaling in Eq.~\eqref{eq:time_delay_21_phys}, this small relative deviation translates into a significantly larger absolute correction for M87*, reaching up to $\sim 20~\mathrm{min}$ for the $2S$ and $4S$ configurations. In contrast, for Sgr A*, the corresponding deviations remain at the level of $10^{-2}~\mathrm{min}$.

\begin{figure*}[t]
    \centering
    \includegraphics[width=8cm]{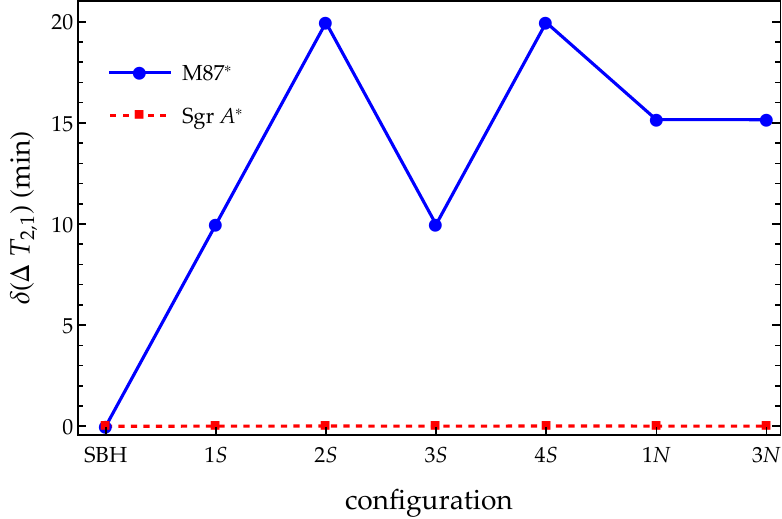} \qquad
    \includegraphics[width=8cm]{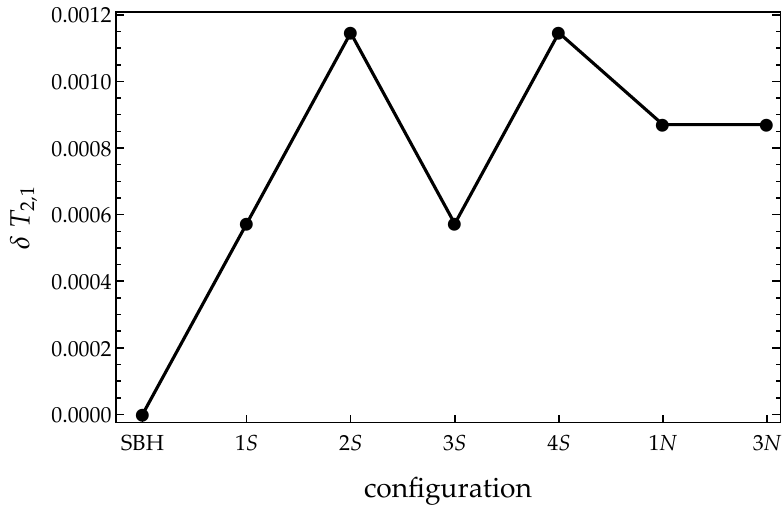}
    \caption{Halo-induced correction to the time delay between the first two relativistic images. Left: absolute deviation $\delta(\Delta T_{2,1})$ in minutes for M87* and Sgr A*. Right: relative deviation $\delta T_{2,1}$, which follows the fractional change of the critical impact parameter.}
    \label{fig:time_delay}
\end{figure*}

To visualize these trends more clearly, we show in Fig.~\ref{fig:time_delay} the absolute correction and the relative deviation of the time delay for all configurations.
The left panel of Fig.~\ref{fig:time_delay} shows that the absolute correction is much larger for M87* than for Sgr A*. This is not due to a different fractional halo effect, but follows directly from the mass scaling in Eq.~\eqref{eq:time_delay_21_phys}. The mass-normalized quantity $c^3\Delta T_{2,1}^{\rm phys}/(G M_{\rm BH})=2\pi b_c$, or equivalently the relative deviation $\delta T_{2,1}$, isolates the halo contribution through the critical impact parameter. This is shown in the right panel and also in the last column of Table~\ref{tab:time_delay}. Therefore, although the fractional correction is small, the absolute time-delay shift can become more relevant for very massive black holes. Among the observables considered in this work, the time delay provides the clearest amplification of the small differences induced by the halo configurations.


\subsection{Observational degeneracy and hierarchy of lensing signatures}\label{subsec:discussion}

The numerical results obtained in the previous subsections show a clear and consistent pattern. The dark matter halo changes the optical geometry in a systematic way, but the effect is very small for the configurations considered here. This can already be seen from the behavior of $U(r)$, where the halo profiles lie slightly above the Schwarzschild one. The same trend appears in the deflection angle, the shadow radius, the finite-order image positions, the magnification, and finally in the time delay. Therefore, the halo does not produce a random correction. It produces a coherent correction, but this correction is strongly suppressed near the photon sphere.

The reason is that the observables studied here are mainly controlled by a small set of strong-lensing quantities, namely $r_p$, $b_c$, $\ba$, $\bar{\xi}$, and $\epsilon_n$. Among them, the photon-sphere radius $r_p$ is almost unchanged for all the halo models, as shown in Table~\ref{tab:rpbc}. The critical impact parameter $b_c$ is more sensitive, but still changes only at the level of $\mathcal{O}(10^{-3})$. This explains why the shadow radius and the RERs show visible but very small shifts, while remaining well inside the present observational bounds for Sgr A*. In this sense, the shadow size alone cannot distinguish the halo configurations with current EHT-level uncertainties.

The finite-order image positions provide a slightly more detailed view of the same behavior. The deviations $\delta\theta_1$ and $\delta\theta_2$ follow almost the same pattern, which means that the halo mainly shifts the relativistic image sequence as a whole, rather than changing its internal spacing in a strong way. The separations $\Delta\theta_{12}$ and $\Delta\theta_{23}$ also remain very small. Therefore, these quantities reveal the presence of a systematic halo correction, but they do not fully break the degeneracy among the models.

A similar conclusion is obtained from the magnification analysis. The relative magnification deviations $\delta\mu_1$ and $\delta\mu_2$ are practically the same for each halo configuration. This means that the first two magnifications are almost rescaled by the same factor. As a consequence, the magnitude difference $\Delta m_{12}$ remains unchanged. The logarithm in the definition of $\Delta m_{12}$ also makes any residual difference even less visible. Hence, the magnification confirms the small halo imprint, but it does not provide a strong discriminator in the present setup.

The ordering of the deviations also has a simple physical interpretation. The configurations $2S$ and $4S$ usually give the largest corrections, while $1S$ and $3S$ are almost degenerate. This is not only controlled by the total halo mass, but rather by the effective compactness of the halo in the region probed by the photon trajectories. A more extended halo can have a larger total mass but still produces a weaker local correction near the photon sphere. This explains why some models with different total masses behave almost in the same way, and why the pairs $1S$--$3S$, $2S$--$4S$, and $1N$--$3N$ appear nearly degenerate in several observables.

Among all quantities considered here, the time delay gives the clearest amplification of the small halo-induced differences. The relative deviation $\delta T_{2,1}$ is again of order $\mathcal{O}(10^{-3})$, because at leading order it is directly related to the fractional change of $b_c$. However, the corresponding absolute correction scales with the black hole mass. For M87*, the $2S$ and $4S$ configurations can produce shifts close to $20~\mathrm{min}$, while for Sgr A* the same relative correction remains at the level of $10^{-2}~\mathrm{min}$. Therefore, the mass-normalized quantity $c^3\Delta T_{2,1}^{\rm phys}/(G M_{\rm BH})=2\pi b_c$ is the clean diagnostic of the halo geometry, while the physical time delay is more promising for very massive black holes.

The main result of this numerical study is, hence, that the considered black hole--dark matter configurations are strongly degenerate with the Schwarzschild case under standard strong-lensing observables. This is not a negative result. It rather shows that these halo profiles can remain hidden in the usual shadow and image-position signatures, even though they leave a small and ordered imprint in the lensing quantities. This provides a useful baseline for the next step of the work, where a perturbative treatment of the black hole effect on the halo profile can be included to see whether the degeneracy is weakened or amplified.

\section{Conclusions}\label{sec:conclusions}

In this work, we have studied the strong gravitational lensing properties of black holes embedded in self-interacting scalar field dark matter halos, together with NFW-type configurations for comparison. Using the Einstein cluster formalism, we reconstructed the corresponding spacetime geometries and investigated how the surrounding halo affects the propagation of photons near the black hole.

We first analyzed the effective function $U(r)$ and the corresponding photon sphere properties. Our results show that the photon sphere radius $r_p$ remains almost unchanged for all halo configurations considered here, while the critical impact parameter $b_c$ receives slightly larger corrections. This indicates that the near-photon-sphere geometry remains very close to the vacuum Schwarzschild solution, even in the presence of comparatively massive dark matter halos.

We then studied different strong-lensing observables, including relativistic Einstein rings, finite-order image positions, image separations, magnifications, and time delays, with particular attention to M87* and Sgr A*. In general, the dark matter halos considered in this work produce only small corrections relative to the Schwarzschild case, typically at the level of $\mathcal{O}(10^{-3})$ or smaller. In particular, the finite-order image positions and magnification hierarchy show a strong observational degeneracy among the different halo models.

Despite this degeneracy, some systematic differences remain visible. Among the observables considered here, the time delay between relativistic images provides the clearest amplification of the halo-induced corrections, especially for very massive black holes such as M87*. This suggests that time-domain strong-lensing observables may provide a more promising avenue to probe the effects of dark matter distributions around black holes than purely angular observables.

Finally, the present analysis adopts an unperturbed dark matter profile around the black hole as a simplified framework to isolate the dominant lensing effects. A more complete treatment could include the gravitational influence of the black hole on the halo distribution itself, potentially modifying the inner density profile and the corresponding strong-lensing observables. Nevertheless, the present results provide a first step toward characterizing how realistic dark matter environments can imprint subtle but potentially observable signatures on strong gravitational lensing around black holes, especially in view of future high-precision measurements of black hole shadows with next-generation facilities such as the ngEHT \cite{Johnson:2023ynn}.

\section*{Acknowledgements}
M.F. acknowledges financial support from Agencia Nacional de Investigación y Desarrollo (ANID) through the FONDECYT postdoctoral Grant No.~3260029. The authors also acknowledge the use of AI tools only for language polishing and improving the clarity of the manuscript.

\bibliographystyle{ieeetr}
\bibliography{biblio_v1.bib}

\end{document}